\documentclass{iopart}             
\usepackage[T1]{fontenc}
\usepackage{graphicx}
\usepackage{url,graphics,epsfig}
\usepackage{amssymb}

\usepackage{tikz}

\newcommand*\mycirc[1]{%
  \begin{tikzpicture}[baseline=(C.base)]
    \node[draw,circle,inner sep=1pt](C) {#1};
  \end{tikzpicture}}

\begin{document}

\jl{2}
%
%
%
\def\etal{{\it et al~}}
%
%
%
%
%
%
\setlength{\arraycolsep}{2.5pt}             

\title[{{\it K}-shell Photoionization of  Be-like and Li-like Ions of Atomic Nitrogen}]{{\it K}-Shell Photoionization 
	of  Be-like and Li-like Ions of Atomic Nitrogen: Experiment and Theory}

\author{  M M Al Shorman$^{1}$, M F Gharaibeh$^{2}$,
              J M Bizau$^{1,3}\footnote[1]{Corresponding author, E-mail: jean-marc.bizau@u-psud.fr}$,  D Cubaynes$^{1,3}$,
              S Guilbaud$^{1}$,  N El Hassan$^{1}\footnote[2]{Present address: Laboratoire Structures, PropriŽtŽs et ModŽlisation 
              des Solides (SPMS) UMR CNRS 8580, Ecole Centrale Paris, 1, Grande Voie des Vignes, 92295 Ch‰tenay-Malabry, France}$,  
              C Miron$^{3}$, C Nicolas$^{3}$, E Robert$^{3}$,  I Sakho$^{4}$,
              C Blancard$^{5}$ and B M McLaughlin$^{6,7}\footnote[3]{Corresponding author, E-mail: b.mclaughlin@qub.ac.uk}$}

\address{$^{1}$Institut des Sciences Mol\'{e}culaires d'Orsay (ISMO), CNRS UMR 8214, 
			Universit\'{e} Paris-Sud, B\^{a}t. 350, F-91405 Orsay cedex, France}

\address{$^{2}$Department of Physics, Jordan University of Science and Technology, Irbid 22110, Jordan}

\address{$^{3}$Synchrotron SOLEIL - L'Orme des Merisiers, Saint-Aubin - BP 48 91192 Gif-sur-Yvette cedex, France}

\address{$^{4}$Department of Physics, UFR of Sciences and Technologies, 
                           University Assane Seck of Ziguinchor, Ziguinchor, Senegal}

\address{$^{5}$CEA-DAM-DIF, Bruy$\grave{\rm e}$res-le-Ch\^{a}tel, F-91297 Arpajon Cedex, France}

\address{$^{6}$Centre for Theoretical Atomic, Molecular and Optical Physics (CTAMOP),\\
			School of Mathematics and Physics, Queen's University Belfast, 
			Belfast BT7 1NN, Northern Ireland, UK}

\address{$^{7}$Institute for Theoretical Atomic and Molecular Physics (ITAMP),\\
			Harvard Smithsonian Center for Astrophysics, MS-14, Cambridge, MA 02138, USA}


%
%

\begin{abstract}
Absolute cross sections for the {\it K}-shell photoionization 
of Be-like and Li-like  atomic nitrogen ions were measured by
employing the ion-photon merged-beam technique at the
SOLEIL synchrotron radiation facility in Saint-Aubin, France.
High-resolution spectroscopy at nominal resolutions 
of 38 meV, 56 meV,  111 meV, 133 meV FWHM 
for Be-like and 125 meV FWHM for Li-like atomic nitrogen ions
 was achieved for the photon energies ranging from 410 eV up to 460 eV. 
 The experimental measurements are compared  with theoretical estimates
from the multi-configuration Dirac-Fock (MCDF), 
R-matrix  and an empirical method.
The interplay between experiment  
and theory enabled the identification and 
characterisation of the strong 1s  $\rightarrow$ 2p 
resonances features observed in the {\it K}-shell spectra of each ion and the region 
around 460 eV for the 1s  $\rightarrow$ 3p resonance of the N$^{3+}$ ion
yielding suitable agreement with experiment. 
\end{abstract}

%
%

\pacs{32.80.Fb, 31.15.Ar, 32.80.Hd, and 32.70.-n}


\vspace{1.0cm}
\begin{flushleft}
Short title: {\it K}-shell photoionization of Be-like and Li-like atomic nitrogen ions\\
\vspace{1cm} Draft for J. Phys. B: At. Mol. \& Opt. Phys: \today
\end{flushleft}

\maketitle
%
%
%
\section{Introduction}
Photoabsorption (PA) and Photoionization (PI) are fundamental atomic processes that play important
roles in many physical systems, including a broad range
of astrophysical objects as diverse as quasi-stellar objects, the atmosphere of
hot stars, protoplanetary nebula, HII regions, novae and supernovae.
Satellites {\it Chandra} and {\it XMM-Newton} currently provide a wealth of x-ray spectra of 
astronomical objects; the lack of high-quality atomic data hammers
the interpretation of these spectra \cite{McLaughlin2001, Brickhouse2010, Kallman2010, Quinet2011,McLaughlin2013}. 
Studies recently carried out on carbon and its ions showed it  was
necessary to have good quality data to model the observations in the x-ray spectrum of the bright blazar
Mkn 421 observed by the Chandra LETG+HRC-S \cite{McLaughlin2010}.
Spectroscopy in the soft x-ray region (5-45 \AA) including 
{\it K}-shell transitions of C, N, O, Ne, S and Si, in neutral, and low stages of ionisation, 
L-shell transitions of Fe and Ni, provide a valuable tool for investigating the extreme 
environments in active galactic nuclei (AGN's),
binary systems, cataclysmic variable stars (CV's) and Wolf-Rayet Stars \cite{Skinner2010}
as well as the interstellar media (ISM) \cite{Garcia2011}. 
One may expect similar results concerning the chemical composition of the ISM  
if  accurate data are available on neutral \cite{Witthoeft2009,McLaughlin2011,Brickhouse2010,McLaughlin2013}, 
and various charge stages of ionisation of atomic nitrogen {\it K}-edge cross sections \cite{Witthoeft2009,Soleil2011}. 
Photoionization models of the brightest knot of star formation in the blue
compact dwarf galaxy Mrk 209 required abundances for ions of oxygen and nitrogen \cite{Diaz2007}.
Therefore PI cross section data and abundances for carbon, nitrogen, and oxygen in their various stages  of ionization are 
essential for photoionization models applied to the plasma modelling in a variety of planetary nebulae \cite{Bohigas2008}. 
Nitrogen abundance in particular plays a fundamental role in $\eta$ Car studies, because it is a key tracer
of CNO processing \cite{Verner2005}.  Li-like atomic nitrogen ions (N V)
 are used in the determination of photoionization structures of pressure-supported gas clouds in gravitationally
dominant dark matter mini-halos in the extended Galactic halo or Local Group environment \cite{Sternberg2004} with the
 modelling code CLOUDY \cite{Ferland1998,Ferland2003}.

X-ray  spectra obtained by {\it Chandra} from sources such as Capella, Procyon, and HR 1099 are
used as standards to benchmark plasma spectral modelling codes.
He-like nitrogen (N VI) lines have been observed with {\it Chandra} and {\it XMM-Newton} in the x-ray spectra 
of Capella and Procyon \cite{Mewe2001,Ness2001}, the M dwarf binary YY Gem \cite{stelzer2002}, and the recorded outburst of
 the recurrent nova RS Oph \cite{Ness2007},  in the wavelengths region 28.7 \AA -- 29.6 \AA ~(420 eV)  that
are attributed to 1s $\rightarrow$ 2$\ell$ transitions. 
{\it XMM-Newton} observations of the fast classical nova V2491 Cyg \cite{Ness2011} have also indicated the 
N VI, K$_{\alpha}$  and K$_{\beta}$ lines are present at about 28.78 \AA~ and 24.90 \AA. 
The He-like series lines of N VI and O VII are detected up to 1s  $\rightarrow$ 5p, and
for O VII, the recombination/ionization continuum at 16.77 \AA~(739.3 eV) is present between 16.6 \AA -- 16.8 \AA. He-like N VI has also been observed 
in the {\it Chandra}-LETGS x-ray spectroscopy of NGC 5548  \cite{Kaastra2002}.

In the x-ray  community, Electron-Beam-Ion-Trap (EBIT) measurements (used for calibrating resonance energies), 
have been carried out for the inner-shell 1s $\rightarrow$ 2$\ell$ transitions in He-like and Li-like nitrogen ions \cite{EBIT1999} 
(used in plasma modelling to determine impurity transport properties \cite{Rice1997}). 
He-like and Li-like x-ray lines  energies of atomic nitrogen have been measured at the 
Lawrence Livermore  National Laboratory (LLNL) EBIT \cite{EBIT1999}.
In EBIT experiments, the spectrum is contaminated and blended with ions in multiple stages of ionization, 
making spectral interpretation fraught with difficulties, unlike the cleaner higher-resolution spectra obtained from third generation 
synchrotron radiation facilities such at the Advanced Light Source, BESSY II, SOLEIL, ASTRID II and Petra III. 

Photoionization cross sections used for  the modelling of astrophysical
phenomena have mainly been provided by theoretical methods,
due to limited experimental data being available. As a consequence,
significant effort has been put into improving the
quality of calculated data using state-of-the-art theoretical methods. 
Recent advances in the determination of atomic parameters for modeling {\it K} lines
in cosmically abundant elements have been reviewed by Quinet and co-workers \cite{Quinet2011}.
Until recently, many of these calculations have
not been severely tested by experiment, and this remains an urgent
task \cite{bizau2005}. Experimental {\it K}-shell photoionization cross section measurements 
have been made by various groups on a variety of atoms and ions of astrophysical interest;
He-like Li$^{+}$ \cite{Scully2006,Scully2007,DPI2013},  Li atoms \cite{Diehl1996},
Li-like  B$^{2+}$ \cite{Mueller2010}, C$^{3+}$  \cite{Mueller2009}, 
Be-like B$^{+}$ \cite{Mueller2013} , C$^{2+}$ \cite{Scully2005}, 
B-like C$^{+}$ \cite{Schlachter2004},
C-like N$^{+}$ \cite{Soleil2011},
N-like O$^{+}$ \cite{Kawatsura2002},
F-like Ne$^{+}$ \cite{Yamaoka2001},
neutral nitrogen \cite{McLaughlin2011} 
and oxygen \cite{Krause1994,Menzel1996,Stolte1997,Stolte2013}, 
valence shell studies on B-like ions, N$^{2+}$, O$^{3+}$ and F$^{4+}$ \cite{bizau2005}. 
C$\ell$-like Ar$^{+}$ \cite{Phaneuf2011}, 
Mg-like Fe$^{14+}$ \cite{Simon2010}, 
As-like Se$^{+}$ \cite{Ballance2012,Esteves2011,Sterling2011} and
Br-like Kr$^{+}$ \cite{Bizau2011,McLaughlin2012,Hino2012}.  
All this experimental data has been compared with
various modern theoretical methods.

Theoretical {\it K}-shell photoionization cross sections for the isonuclear 
C~I -- C~IV ions assembled recently to model the {\it Chandra} x-ray absorption 
spectrum of the blazar Mkn 421  \cite{McLaughlin2010} were found to be in excellent agreement 
with the astrophysical  observations. Additionally, {\it K}-shell photoionization cross sections calculations on neutral nitrogen 
showed excellent accord with high resolution measurements made at the Advanced Light Source 
radiation facility in Berkeley, California \cite{McLaughlin2011} as have similar cross section calculations on singly and doubly ionised atomic 
nitrogen when compared with measurements from the SOLEIL synchrotron facility in Saint--Aubin, France \cite{Soleil2011,Soleil2013}. 
 In fact, the majority of the high-resolution experimental 
studies from the third generation light sources have been shown to be in excellent agreement with detailed theoretical 
calculations performed  using the state-of-the-art R-matrix method \cite{rmat,codes}  and with other theoretical approaches.

The present  experimental and theoretical work on the prototype Be-like and Li-like atomic nitrogen ions provides 
cross section data for the photoionization of x-rays in the vicinity of the {\it K}-edge, 
where strong n=2 inner-shell resonance states are observed. 
This work follows our earlier successful work on 
{\it K}-shell investigations for singly and doubly ionised atomic nitrogen \cite{Soleil2011,Soleil2013}.
To our knowledge, there would appear to be limited experimental studies for resonance Auger energies reported  to date
on either  Be-like or Li-like atomic nitrogen ions  \cite{EBIT1999,Tondello1977,Niehaus1987} 
for photon energies in the vicinity of the  {\it K}-edge region.  
For Be-like atomic nitrogen, previous experimental and theoretical 
studies have been made in the  valence region and in the near threshold region 
\cite{bizau2005,Mueller2007,Mueller2010b,Simon2010}
where it was necessary to include both the ground state and metastable excited states 
in the theoretical work in order to achieve suitable agreement with experiment. 
We follow a similar prescription here in the vicinity of the {\it K}-shell 
energy region as  the Be-like (N$^{3+}$) atomic nitrogen  ions produced in the SOLEIL synchrotron 
radiation experiments are not purely in their ground state.  
Theoretical studies are an essential ingredient in order to determine the metastable constituents in the beam.
In the case of Li-like atomic nitrogen ions one only needs to consider the case of the 
ground state as no metastable states are present in the parent beam.

{\it K}-shell photoionization when followed by Auger decay couples
three or more ionization stages rather than two in 
the usual equations of ionization equilibrium \cite{Petrini1997}.
 The 1s $\rightarrow$ np photo-excitation processes involved in the interaction of a photon
with the $\rm 1s^22s^2~^1S$ ground-state of the Be-like nitrogen ion is;
$$
 h\nu + {\rm N^{3+}(1s^22s^2~^1S)}  \rightarrow  {\rm N^{3+} ~ (1s2s^2np[^{1}P^{\circ}] ) }
 $$
 $$
 \swarrow \quad \searrow
 $$
 $$
{\rm  N^{4+}~ (1s^22s~^2S) + e^- ({\it k^2_{\ell}})} {\quad} {\rm or} {\quad} {\rm  N^{4+}~ (1s^2np~^2P) + e^- ({\it k^2_{\ell}}),} 
$$
n=2,3, and where $k^2_{\ell}$ is the outgoing energy of the continuum electron with angular momentum $\ell$. 
The strongest Auger decay channels being the spectator KLL channels where the Rydberg, np electron does not 
participate in the Auger decay.

Experimental studies on Be-like atomic nitrogen ions, in their ground state
$\rm 1s^22s^2~^1S$, are contaminated by the presence of 
metastable states. In the SOLEIL experiments, N$^{3+}$ ions 
are produced in the gas-phase from an Electron-Cyclotron-Resonance-Ion-Source (ECRIS) 
therefore metastable states $\rm 1s^22s2p~^3P^{\circ}$  may be present 
in the parent ion beam.  The $\rm 1s^22s2p~^3P^{\circ}$ metastable 
state, autoionization processes from the 1s $\rightarrow$ 2p photo-excitation process  are;
$$
 h\nu + {\rm N^{3+}(1s^22s2p~^3P^{\circ})}
 $$
$$
\downarrow
$$
$$
{\rm  N^{3+} (1s2s[^{1,3}S]2p^2(^3P,^1D,^1S)]^3S,^3P,^3D)}
$$
$$
\swarrow \quad \searrow
$$
$$
{\rm  N^{4+}~ (1s^22s~^2S) + e^- ({\it k^2_{\ell}})} {\quad} {\rm or} {\quad} {\rm  N^{4+}~ (1s^22p~^2P) + e^- ({\it k^2_{\ell}}).} 
$$
For the case of Li-like ions, the strongest excitation processes in the interaction of a photon
with the $\rm 1s^22s~^2S_{1/2}$ ground-state of the Li-like nitrogen
ion is the 1s $\rightarrow$ 2p photo-excitation process;
$$
 h\nu + {\rm N^{4+}(1s^22s~^2S_{1/2}})  \rightarrow  {\rm N^{4+} ~ (1s[2s2p~^{1,3}P]~^2P^{\circ}_{1/2,3/2}) }
 $$
 $$
 \downarrow
 $$
 $$
{\rm  N^{5+}~ (1s^2~^1S_0) + e^- ({\it k^2_{\ell}}),}
$$
Detailed experimental photoionization measurements have been performed 
for the first time on these prototype Be-like and Li-like systems in the photon energy region of the {\it K}-edge.

In our previous publication on {\it K}-shell photoionisation of N$^+$ ions \cite{Soleil2011}
we high-lighted  the limitations of central fields calculations using Harte-Slater of Dirac-Slater potentials 
\cite{Slater1960,HS1963,Reilman1979,Band1979,Verner1993}.
We stress again that results determined from these methods used in spectral 
modelling should be treated with caution.  

For Be-like atomic nitrogen ions, state-of-the-art  {\it ab initio} calculations for Auger inner-shell processes were carried out by
Petrini and de Ara\'ujo \cite{Petrini1997} and by Berrington and co-workers \cite{Berrington1997} 
using the R-matrix method \cite{rmat} and followed a 
similar procedure to the work on {\it K}-shell studies for the Be-like B$^+$ ion \cite{Petrini1981}. 
Petrini and co-workers \cite{Petrini1998} noted that once  the 1s-hole was created in the ions, by single photoionization, with
simultaneous shake-up and shake-off processes, Auger decay
populates directly excited states of the residual ions, which then produces UV lines.
Similarly, for Li-like atomic nitrogen ions, Charro and co-workers \cite{Charro2000} performed R-matrix calculations in
$LS$-coupling for photoionization cross sections and compared and contrasted their results with 
previous central field and R-matrix methods as only limited experimental data was available. 
However, no attempt was made in their calculations at determining resonance parameters.

About a decade later this work was further extended by  Garcia and co-workers \cite{Witthoeft2009}, 
using the optical potential method within the Breit-Pauli R-matrix formalism \cite{rmat,codes,damp,Burke2011}.  
Photoionization from the ground state, along the nitrogen iso-nuclear sequence was investigated, 
in the photon energy region of the {\it K}-edge.  
We note that for {\it K}-shell photoionization, the cross section calculations of 
Garcia and co-workers \cite{Witthoeft2009}  included both radiation and Auger
damping, which cause the smearing of the {\it K} edge.
Due to the lack of experimental data  being available at that time 
a comparison with previous theoretical results (of lower calibre) 
 were only possible. Garcia and co-workers  \cite{Witthoeft2009}  pointed out
 these earlier central field calculations do not take account of the resonance features 
 that dominate the cross sections near the {\it K}-edge.
In the present study we compare our  theoretical results
from the multi-configuration Dirac Fock (MCDF) and the R-matrix with pseudo states methods (RMPS),
an empirical  fitting approximation \cite{Sakho2011} 
(derived from previous experimental measurements on resonance energies 
and Auger widths of the iso-electronic sequence),
prior theoretical results \cite{Chen1983,Chen1985,Witthoeft2009} and 
current experimental measurements made at the SOLEIL synchrotron radiation facility.

In this paper detailed theoretical calculations and experimental measurements 
are presented for the  {\it K}-shell  single photon ionisation cross sections 
of Be-like atomic nitrogen ions (410 eV -- 415 eV and 460 -- 460.4 eV)
and Li-like atomic nitrogen ions, (420 eV -- 426 eV). Our theoretical 
predictions from the MCDF and R-matrix methods enabled identification of 
the $\rm 1s \rightarrow 2p$  resonances and their parameters, 
observed in the Be-like and Li-like nitrogen spectra and
the $\rm 1s \rightarrow 3p$ resonance around 460 eV in Be-like nitrogen.
The present investigation provides absolute values (experimental and theoretical)
for cross sections along with the n=2  inner-shell resonance energies, natural line-widths 
and resonance strengths, for a photon colliding with the
$\rm 1s^22s^2~^1S$,  and $\rm 1s^22s2p~^3P^{\circ}$ states of the N$^{3+}$ ion.  
For Li-like ions, only the ground state $\rm 1s^22s~^2S_{1/2}$ is present in the beam. 

The layout of this paper is as follows. Section 2 presents the experimental procedure used. 
Section 3 gives an overview of the theoretical work. Section 4 compares and contrasts 
the results obtained from the experimental and theoretical methods.
Finally in section 5 conclusions are drawn from the present investigation.

\section{Experiment}\label{sec:exp}

Cross sections for photoionization of Be-like and Li-like atomic nitrogen 
ions were measured in the range where {\it K}-shell photoionization
occurs. The experiment was performed at the MAIA (Multi-Analysis Ion Apparatus)
set-up, permanently installed on branch A of the PLEIADES beam line  \cite{Pleiades2010,Miron2013} at SOLEIL,
the French National Synchrotron Radiation Facility, located in Saint--Aubin,
France. Details of the experimental setup were outlined in a previous 
publication on N$^{+}$ \cite{Soleil2011}. 
The N$^{3+}$ and N$^{4+}$ ions are produced in a permanent magnet
electron cyclotron resonance ion source (ECRIS). 
Collimated N$^{3+}$ and N$^{4+}$ ion-beam currents
up to 100 nA were extracted from the ion source after biasing the ion source by
+2 kV and then selected by mass per charge ratio using a dipole magnet selector.
The ion beam was placed on the same axis as the photon beam by using
electrostatic deflectors and einzel lenses to focus the beam. 
After the interaction region between the photon and the ion beams, another dipole magnet
separates the primary beam and the beam of ions which have gained one 
(or several) charge(s) in the interaction, the so-called photo-ions. The
primary ions are collected in a Faraday cup and the photon-ions are detected by
multi-channel plates detector. The photon current is measured by a calibrated
photodiode. Ions with the same charge as the photo-ions can also 
be produced by collisions between the primary
ions and the residual gas or stripping on the walls in the interaction region.
This background signal is subtracted by chopping the photon beam, collecting the
data with and without photons for 20 seconds accumulation time.

%
%

\begin{table}
\caption{\label{expt} Experimental parameters used for evaluating the absolute
			       cross section for N$^{3+}$ ions measured at  a photon energy of 412.5 eV.}
\begin{indented}
\item[]\begin{tabular}{@{}*{7}{l}}
\br
{\it Signal}			&340 Hz						\\
Noise			&250 Hz						\\
Velocity			&2.0 $\times$ 10$^{5}$ ms$^{-1}$	\\
Photon flux		&3.0 $\times$ 10$^{11}$ s$^{-1}$	\\
Ion current		&100 nA						\\
Detector efficiency	&0.36						\\
Form factor		&59							\\
\br
\end{tabular}
\end{indented}
\end{table}

For the absolute measurements of the PI cross sections, a -1000 V bias is applied
on the 50 cm long interaction-region and the data are collected with 30 meV
photon energy steps. The overlap of the two beams and the density distributions
of the interacted particles is determined in three dimensions by using two sets
of three scanning slits.  The cross sections obtained have an estimated systematic uncertainty of 15\%. 
In another spectroscopy mode, no bias is applied to the
interaction region allowing the photon and ion beams to interact over about 1 m
and to scan the photon energy with a finer step. 
In this mode, only relative cross sections can be measured. 
They are later normalised on the cross sections determined 
in the absolute mode assuming the area under the resonances to be the same.
Table \ref{expt} gives typical experimental parameters used to evaluate 
absolute cross sections for N$^{3+}$ ions at  a photon energy of 412.5 eV.
The energy and band width of the photon beam are 
calibrated separately using a gas cell and N$_{2}$ (1s $\rightarrow{}$
$\pi{}$*$_{g}$  v=0) photoionization lines, located at 400.87 eV \cite{Sodhi1984} and 
Ar 2p$_{3/2}$$^{-1}$4s at 244.39 eV \cite{Kato2007}.
The photon energy, once corrected for Doppler shift, has an uncertainty
 of approximately 30 meV.  Outstanding possibilities 
in terms of spectral resolution and flux at the N$_2$ (1s$^{-1}$) K-edge have been discussed 
recently by Miron and co-workers \cite{Miron2012,Kimberg2013}.

\section{Theory}\label{sec:Theory}

\subsection{SCUNC: Li-like and Be-like nitrogen}\label{subsec:SU_Theory}
In the framework of the Screening Constant by Unit Nuclear Charge (SCUNC) 
formalism \cite{Sakho2011,Sakho2012,Sakho2013}, the total energy
of the core-excited states is expressed in the form given by,
\begin{equation}
E \left( N\ell n\ell^{\prime}; ^{2S+1}L^{\pi} \right) 
= -Z^2  \left[  \frac{1}{N^2} 
                          + \frac{1}{n^2}  \left[ 1 
                          -\beta  \left ( N\ell n\ell^{\prime}; ^{2S+1}L^{\pi}; Z  \right)  \right]^2  \right] .
\end{equation}
where $E\left( N\ell n\ell^{\prime}; ^{2S+1}L^{\pi} \right)$ is in Rydberg units. 
In this equation, the principal quantum numbers $N$ and $n$ are respectively for the inner and the
outer electron of the He-like iso-electronic series. The $\beta$-parameters are screening constants by
unit nuclear charge expanded in inverse powers of $Z$ and are given by the expression,
\begin{equation}
\beta \left(  N \ell n\ell{^{\prime}} ; {^{2S+1}} L{^{\pi}} \right) 
            = \sum_{k=1}^{q}  f_k  \left(  \frac{1}{Z}  \right) ^k 
\end{equation}
where $f_k \left( N \ell n\ell^{\prime}; ^{2S+1}L^{\pi}  \right)$ are 
parameters to be evaluated empirically from a previous experimental measurements.  
Similarly, one may get the Auger widths $\Gamma$ in Rydbergs (1 Rydberg = 13.605698 eV)  from the formula

\begin{equation}
\Gamma ({\rm Ry})  = Z^2  \left[   1 -\frac{ f_1}{Z} \times \frac{Z}{Z_0} 
                                                -\frac{f_1}{Z^3} \times \frac{(Z - Z_0)}{Z^2_0} 
                                                - \frac{f_1}{Z^4} \times \frac{(Z - Z_0)}{Z^3_0}  \right] ^2.
\end{equation}
The experiment measurements of M\"uller and co-workers on Be-like carbon and Li-like carbon and boron \cite{Scully2005, Mueller2009,Mueller2010} 
were used to determine all the appropriate empirical parameters. For Be-like carbon, we note the labelling of the $\rm 1s2s2p^2~^3$D 
and $\rm 1s2s2p^2~^3$P states of M\"uller and co-workers \cite{Scully2005} should be reversed as 
pointed out  in our recent calculations on the carbon iso-nuclear sequence \cite{McLaughlin2010}  used to
model the x-ray spectra of the bright Blazar Mkn 421 observed by the Chandra satellite.

\subsection{MCDF: Li-like and Be-like nitrogen}\label{subsec:MCDF_Theory}

Multi-configuration Dirac-Fock (MCDF) calculations were performed based on a full intermediate 
coupling regime in a $jj$-basis using the code developed by Bruneau  \cite{Bruneau1984}. Photoexcitation cross 
sections have been carried out for  both Li-like and Be-like atomic nitrogen ions in the region of their 
respective K-edges. Only electric dipole transitions have been computed using the Babushkin gauge.
For Be-like atomic nitrogen ion the following initial configurations have been considered:
$\rm 1s^2 {\overline{2s}}^2$, $\rm 1s^2 {\overline{2s}}\; {\overline{2p}}$, and $\rm 1s^2 {\overline{2p}}^2$. 
Where the bar over the orbital is to indicate that they are different for the initial and final states.
In order to describe the correlation and relaxation effects, multiple orbitals with the same 
quantum number have been used. Then, the following final configurations have been considered:
 $\rm 1s 2s^2 2p$, $\rm 1s 2s 2p^2$, $\rm 1s 2p^3$, $\rm 1s 2s^2 3p$, $\rm 1s 2s 2p 3p$, and $\rm 1s2p^2 3p$.
Such notation means that radial functions for n=2 orbitals are not the same for initial 
and final configurations. The wavefunctions have been calculated minimizing the following energy functional:
\begin{equation}
E =\frac{ \sum_{\alpha}(2J_{\alpha} + 1) E_{\alpha}}{2 \sum_{\alpha}(2J_{\alpha} + 1)}    + 
       \frac{ \sum_{\beta}(2J_{\beta} + 1) E_{\beta}}{2 \sum_{\beta}(2J_{\beta} + 1)} 
\end{equation}
where $\alpha$ and $\beta$ run over all the initial and final states, respectively. 

Synthetic spectrum has been constructed as a weighted sum of photoexcitation cross sections 
from both the ground state level  $\rm 1s^2 2s^2 (^1S_0)$ and the metastable state levels 
$\rm 1s^2 2s 2p (^3P^{\circ}_{0,1,2})$. Each electric dipole 
transition has been dressed by a Lorenztian profile, assuming a full width at half maximum (FWHM) 
equal respectively to 66 meV and 80meV for  the 1s-2p and 1s-3p transitions in Be-like atomic nitrogen. Such FWHM values have been 
deduced from the MDCF calculations that were performed separately. A mixture of 60 \% ground and 40 \% metastable 
states has been used to fit the experimental results.  In order to compare directly with the Be-like 
experimental data, the synthetic spectrum has been convoluted with a Gaussian profile to simulate the experimental resolution.
A similar procedure has been used for the Li-like atomic nitrogen ion, 
except that  (i) the following configurations have been considered :
$\rm 1s^2 {\overline{2s}}$, $\rm 1s {\overline{2s}} \;{\overline{2p}}$, and $\rm 1s {\overline{2s}} 3p$, and 
(ii) a Lorentzian FWHM equal to 45 meV and 60 meV has been used respectively for the 1s-2p and 1s-3p transitions in Li-like atomic nitrogen. 
It should be pointed out that for the Li-like atomic nitrogen ion, only the ground state contribution 
has been retained to compare with experimental data.

\subsection{R-matrix: Be-like nitrogen}
The $R$-matrix method \cite{rmat,codes,damp,Burke2011},  
using a version of the codes implemented on parallel architectures
 \cite{ballance06,McLaughlin2012,Ballance2012}  determined
the necessary cross sections.  For Be-like ions both the initial $\rm ^1S$ ground state 
and the  $\rm ^3$P$^{\circ}$ metastable states were required.  Cross section calculations were carried 
out in $LS$-coupling with 390-levels  retained in the close-coupling expansion using the R-matrix with 
pseudo states method (RMPS). The Hartree-Fock $\rm 1s$, $\rm 2s$ and $\rm 2p$ tabulated orbitals of Clementi and Roetti
\cite{Clementi1974} were used with n=3 physical and n=4 pseudo orbitals of the residual N$^{4+}$ ion.  
 The n=4 pseudo-orbitals were determined by energy optimization on the 
ground state of the N$^{4+}$ ion, with the atomic structure code CIV3 \cite{Hibbert1975}.  
The n=4 pseudo-orbitals are used to account for core 
relaxation and electron correlation effects, 
in the multi-configuration interaction target wavefunctions.
The  N$^\mathrm{4+}$ residual 390 ion states used
multi-configuration interaction target wave functions. The non-relativistic
$R$-matrix method determined the energies
of the N${^\mathrm{3+}}$ bound states and all the appropriate cross sections.
We determined  PI cross sections  for the $\rm 1s^22s^2$\, $\rm ^1$S ground state and the  
 $\rm 1s^22s2p$\, $\rm ^3$P$^{\circ}$ metastable  state 

\begin{figure}
\begin{center}
\includegraphics[scale=2.5,height=14.0cm,width=16.0cm]{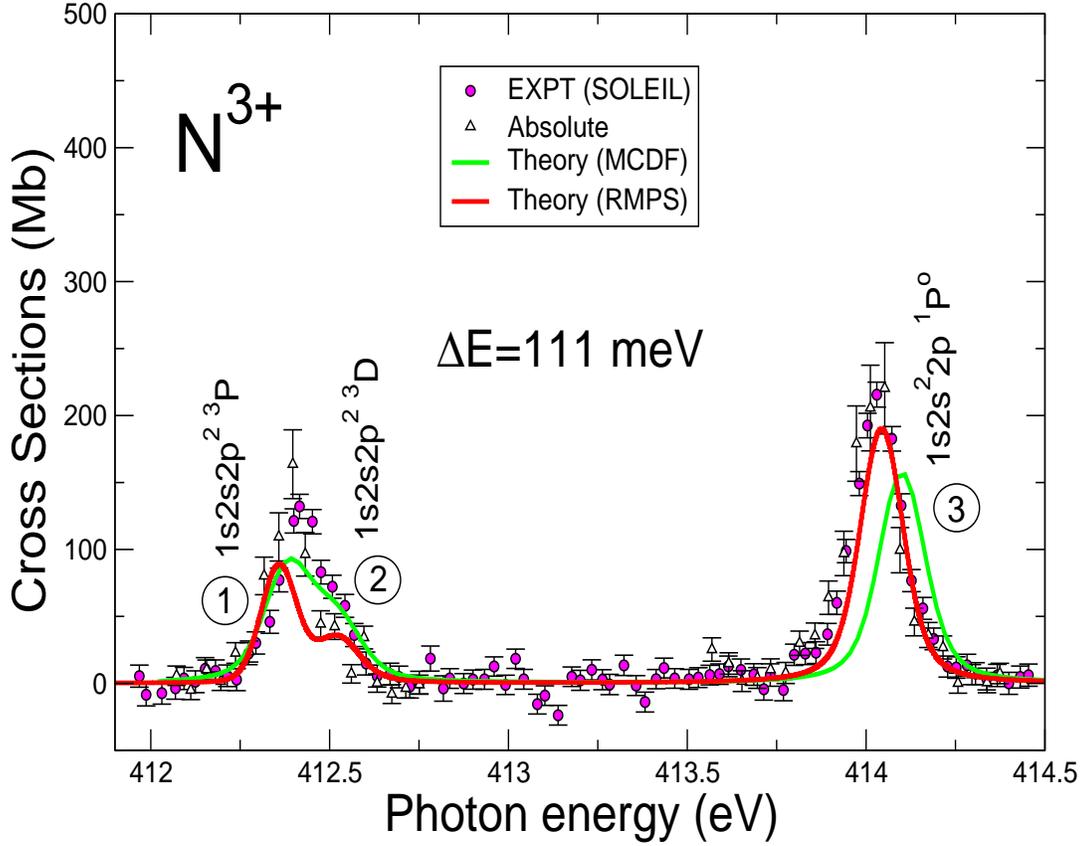}
\caption{\label{fig:111meV}(Colour online) Photoionization cross sections for Be-like atomic nitrogen (N$^{3+}$) ions measured 
								with a 111 meV band pass at the SOLEIL radiation facility. 
								Solid circles :  total photoionization recorded in the relative mode. 
								The error bars represent the statistical uncertainty.
								The absolute measurements (open triangles) total 
								photoionization cross sections have been obtained with a larger energy step. 
								The error bars give the total uncertainty of the experimental data. 
								The MCDF (solid green line) and R-matrix (solid red line) calculations shown are convolution with a Gaussian
								 profile of 111 meV FWHM and an appropriate weighting of the ground and metastable states 
								 (see text for details) to simulate the measurements.  For the metastable $\rm ^3P^{\circ}$ state,
								 the MCDF calculations have been shifted up by +1.46 eV in order to match experiment.
								 Table 2 gives the designation of the resonances~\mycirc{1}  - \mycirc{3} and their parameters.}
\end{center}
\end{figure}

For the electron-ion collision work we
allowed three-electron promotions out of selected base
configurations of N$^\mathrm{3+}$.
The collision work was carried out with twenty
continuum functions and a boundary radius of 8.2 Bohr radii. 
From our RMPS calculations,  the $\rm ^1$S ground state
gave a bound state ionisation potential of  5.69195 Rydbergs 
compared to the experimental value of  5.69420 Rydbergs.
In the case  of the $\rm ^3P^{\circ}$ metastable state,  the ionisation potential from the RMPS
calculations gave 5.08202 Rydbergs, compared to the experimental value of  5.08200 Rydbergs.  
A discrepancy respectively of approximately 31 meV for the ground state and 0.3 meV for the metastable state.
It is seen that both RMPS theoretical results are in excellent agreement with the experimental values
from the NIST tabulations \cite{NIST2012}. 

\begin{figure}
\begin{center}
\includegraphics[scale=2.5,height=14.0cm,width=16.0cm]{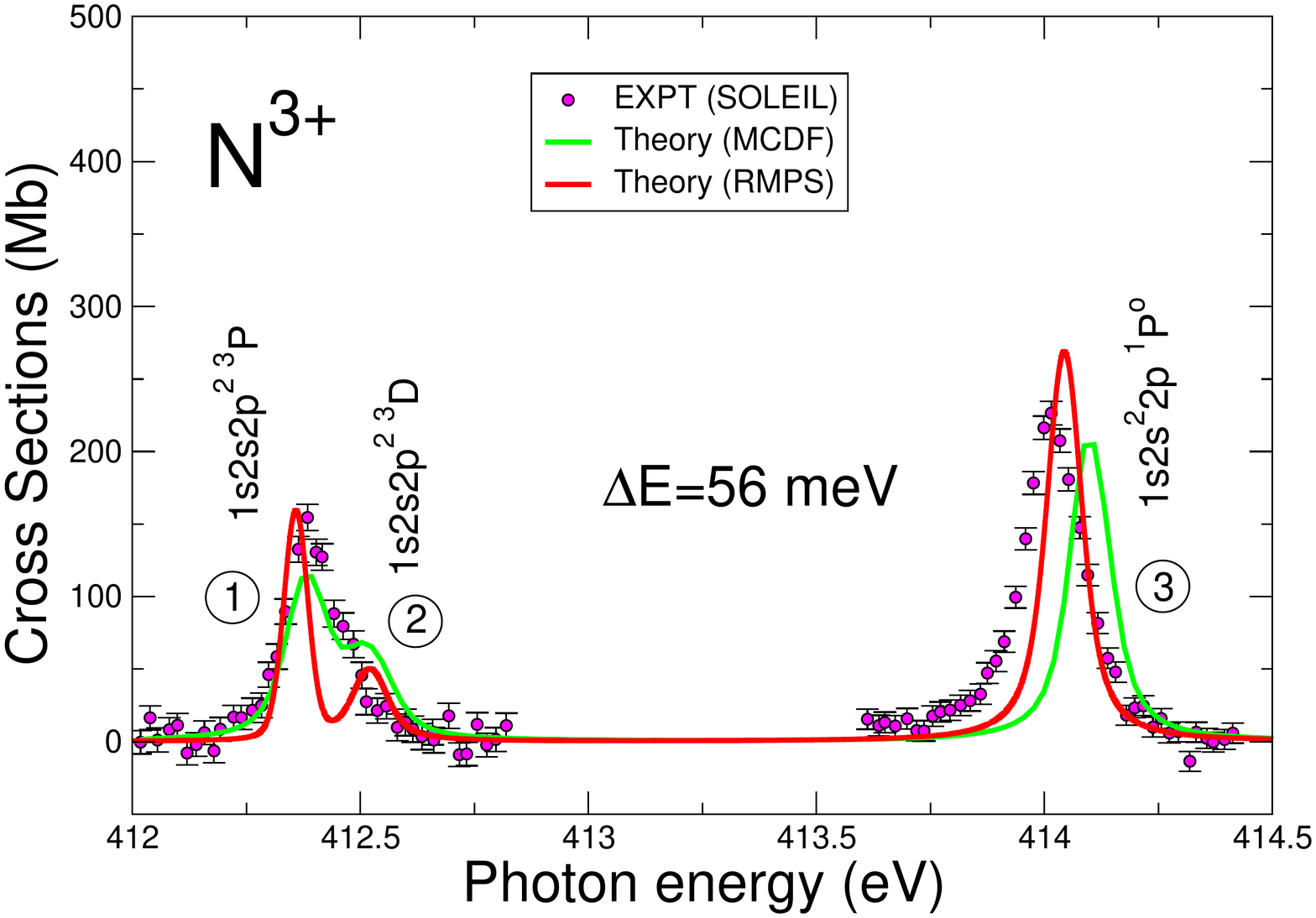}
\caption{\label{fig:56meV}(Colour online) Photoionization cross sections for Be-like atomic nitrogen (N$^{3+}$) ions measured 
								with a 56 meV band pass at the SOLEIL radiation facility. 
								Solid circles :  total photoionization. 
								The error bars give the statistical uncertainty of the experimental data. 
								The MCDF (solid green line) and R-matrix (solid red line) 
								calculations shown are convolution with a Gaussian
								 profile of 56 meV FWHM and an appropriate weighting of the ground and metastable states 
								 (see text for details) to simulate the measurements. For the metastable $\rm ^3P^{\circ}$ state,
								 the MCDF calculations have been shifted up by +1.46 eV in order to match experiment.
								 Table \ref{reson} gives the designation of the resonances~\mycirc{1}  - \mycirc{3} and their parameters.}
\end{center}
\end{figure}

For the $\rm ^1$S ground state and the  $\rm ^3P^{\circ}$
metastable state, the outer region electron-ion collision
problem was solved by selecting an appropriately fine
energy mesh of 2$\times$10$^{-7}$ Rydbergs ($\approx$ 2.72 $\mu$eV)
in order to delineate all the resonance features in the cross sections.  
Radiation and Auger damping were also included in the R-matrix calculations.  

For a direct comparison with the SOLEIL experimental measurements (performed at the various energy resolutions), 
the R-matrix results were convoluted with an appropriate Gaussian function of full-width half-maximum (FWHM)
and an admixture of 60\% ground state  and 40 \% metastable state was used to simulate experiment.
The peaks  found in  the theoretical photoionization cross section 
spectrum were fitted to Fano profiles for overlapping resonances 
\cite{Fano1968,Shore1967,Shore1967b,Shore1968,Ederer1971,Ederer1976,Morgan2008}
instead of the energy derivative of the eigenphase sum method \cite{keith1996,keith1998,keith1999}. 

\begin{figure}
\begin{center}
\includegraphics[scale=2.5,height=14.0cm,width=16.0cm]{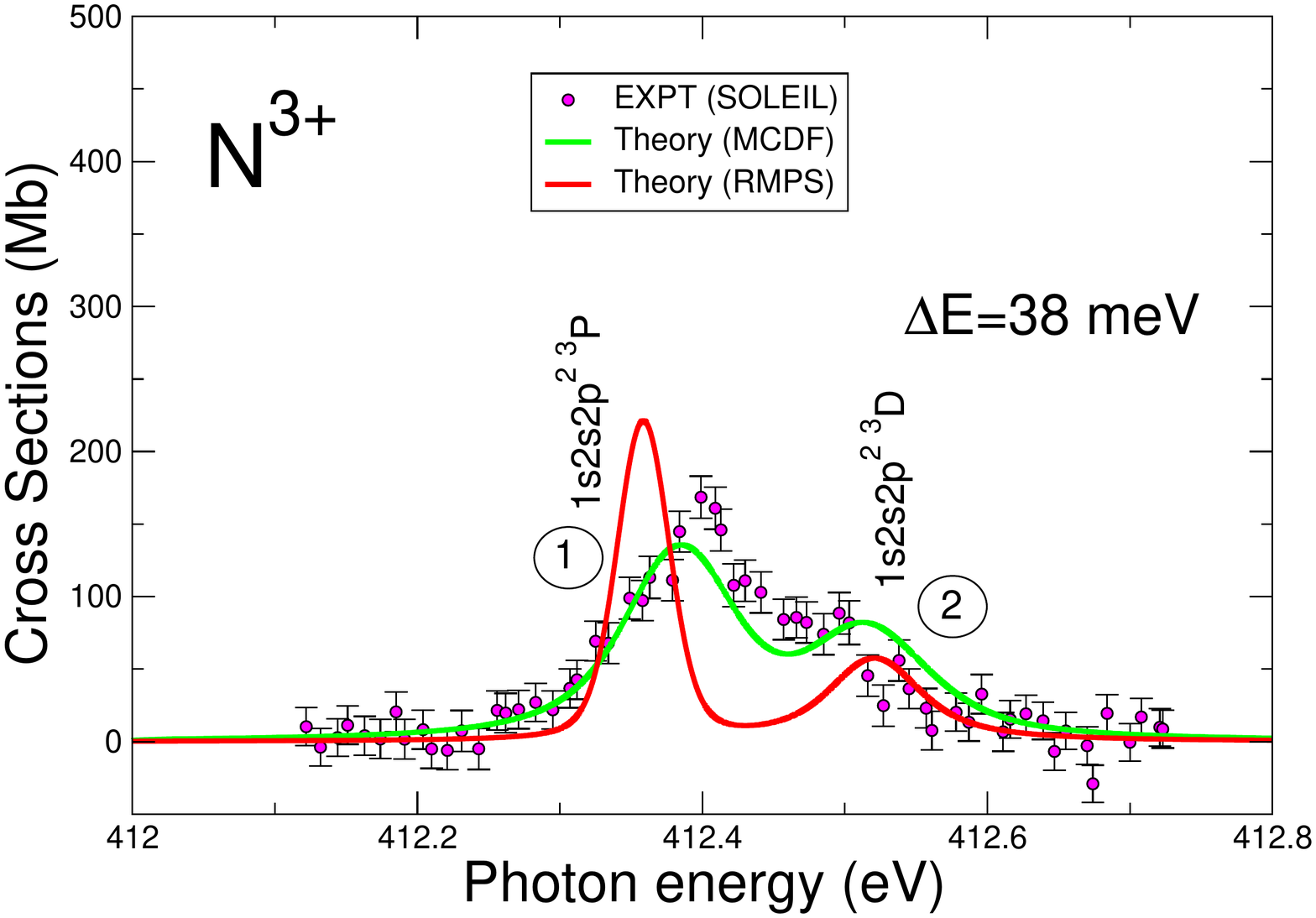}
\caption{\label{fig:38meV}(Colour online) Photoionization cross sections for Be-like atomic nitrogen (N$^{3+}$) ions measured 
								with a 38 meV band pass at the SOLEIL radiation facility. 
								Solid circles :  total photoionization. 
								The error bars give the statistical uncertainty of the experimental data. 
								The MCDF (solid green line) and  R-matrix (solid red line)
								calculations shown are convolution with a Gaussian
								 profile of 38 meV FWHM and an appropriate weighting of the ground state and metastable state 
								 (see text for details) to simulate the measurements.  For the metastable $\rm ^3P^{\circ}$ state,
								 the MCDF calculations have been shifted up by +1.46 eV in order to match experiment.
								 Table \ref{reson} gives the designation of the resonances~\mycirc{1}  - \mycirc{2} and their parameters.}
\end{center}
\end{figure}

%
%
%
%
%
\begin{table}
\caption{\label{reson} Be-like nitrogen (N$^{3+}$), present experimental and theoretical results 
				  for the resonance energies $E_{\rm ph}^{\rm (res)}$ (eV), natural line-widths 
				  $\Gamma$ (meV) and resonance strengths $\overline{\sigma}^{\rm PI}$ (in Mb eV),
           			  for the dominant core photo-excited n=2  states of the N$^{3+}$ ion, in the photon 
			  	  energy region 410 eV to  415 eV compared with previous investigations.  The experimental 
				  error in the calibrated photon energy is estimated to be $\pm$ 30 meV for the 
				  resonance energies. } 
 \lineup
  \begin{tabular}{ccr@{\,}c@{\,}cccc}
\br
 Resonance    		& 					    & \multicolumn{3}{c}{SOLEIL}                                       & \multicolumn{1}{c}{R-matrix} 		& \multicolumn{2}{c}{MCDF/Others}\\
 (Label)            		& 					    & \multicolumn{3}{c}{(Experiment$^{\dagger}$)}      & \multicolumn{1}{c}{(Theory)} 		& \multicolumn{2}{c}{(Theory)}\\
 \ns
 \mr
   $\rm 1s^22s2p\, ^3$P$^{\circ}$ \,$\rightarrow$\, $\rm 1s2s2p^2[^4P]\, ^3$P			
   				& $E_{\rm ph}^{\rm (res)}$     &   		& 412.396 $\pm$ 0.03$^{\dagger}$ & 				& 412.358$^{a}$    	& 	& 410.925$^{b}$\\
  ~\mycirc{1}		&          	 		               &   		&  				 		        &				& 412.026$^{f}$    	&	& 410.130$^{c}$\\
				&          	 		               &   		&  				 		        &				&     				&	& 412.375$^{d}$\\ 
				&          	 		               &   		&  				 		        &				&     				&	& 412.426$^{g}$\\ 
				\\
                                     & $\Gamma$                             & \;\0\0\          & 85 $\pm$ 14             		        &     				&   12$^{a}$  		& 	&   25$^{c}$ \\
		 	         &                                                  &   		&  						        &				&   12$^{f}$    		&	&   11$^{d}$  \\
		 	         &                                                  &   		&  						        &				&     		        		&	&   26$^{g}$ \\
			         \\
				&$\overline{\sigma}^{\rm PI}$&			&  21.08 $\pm$ 4.3		        		&				&  11.18$^{a}$  	&	&   \\
 	   		       \\
  $\rm 1s^22s2p\, ^3$P$^{\circ}$ \,$\rightarrow$\, $\rm 1s2s2p^2[^2D]\, ^3$D			
                                     & $E_{\rm ph}^{\rm (res)}$     &	 		 &412.494 $\pm$ 0.03$^{\dagger}$  & 				& 412.521$^{a}$    	& 	& 411.074$^{b}$\\
  ~ \mycirc{2}	         &           				     &		   	 &  						  	&				& 412.755$^{f}$      	&	& 411.250$^{c}$\\
  			         &           				     &		   	 &  						  	&				&     		         		&	& 412.483	$^{d}$\\
   			         &           				     &		   	 &  						  	&				&     		         		&	& 412.656	$^{g}$\\
 			         &           				     &		   	 &  						  	&				&     		         		&	&   \\
			         \\
 			  	& $\Gamma$ 			     & \;\0\0\    	 & 46  $\pm$ 32     				& 	         			&     59$^{a}$ 		& 	&    22$^{c}$\\
 	    			&           				     &   		 &  						         &				&     63$^{f}$		&	&     58$^{d}$\\ 
  	    			&           				     &   		 &  						         &				&     		 		&	&     62$^{g}$\\ 
				\\  
				&$\overline{\sigma}^{\rm PI}$&			 &  4.70 $\pm$ 2.8		         		&				&     6.4$^{a}$		&	&     \\
	   		       \\
 $\rm 1s^22s^2\,^1$S \,$\rightarrow$\, $\rm 1s2s^22p\, ^1P^{\circ}$				
     				& $E_{\rm ph}^{\rm (res)}$    & 			& 414.033 $\pm$ 0.03$^{\dagger}$&  				& 414.043$^{a}$    	& 	& 414.104	$^{b}$\\
 ~\mycirc{3} 	         &           				    &   		&  							&    	   			& 413.920$^{\ddagger}$&& 412.590$^{c}$\\
 		 	         &           				    &   		&  							&    			 	& 413.872$^{f}$  	&     	& 414.554$^{d}$\\
 		 	         &           				    &   		&  							&    			 	&				&     	& 412.275$^{e}$\\
		 	         &           				    &   		&  							&    			 	&				&     	& 414.290$^{g}$\\
			         \\
				& $\Gamma$			    & \;\0\0\0   	& 93 $\pm$ 13     				&  		     		& 60$^{a}$  		& 	&    27$^{c}$ \\
 	    			&           				   &   			&  							&				& 48$^{f}$     		&	&    58$^{d}$\\   
 	    			&           				   &   			&  							&				&     		 		&	&    65$^{e}$ \\   
 	    			&           				   &   			&  							&				&     		 		&	&    81$^{g}$ \\   
				\\
				&$\overline{\sigma}^{\rm PI}$&			&  42.6 $\pm$ 6.45				&				& 34.3$^{a}$		&	&    \\
				 \\           
 \br
\end{tabular}
~\\
$^{\dagger}$SOLEIL, experimental work.\\
$^{a}$$LS$-coupling, R-matrix   present work,  R-matrix $^{\ddagger}$Berrington and co-workers \cite{Berrington1997}.\\
$^{b}$Multi-Configuration Dirac Fock (MCDF), present work.\\
$^{c}$MCDF, Chen and co-workers \cite{Chen1985}.\\
$^{d}$Saddle-point with complex-rotation method (SPM-CR) \cite{Lin2001,Lin2002}, $^e$(SPM-CR)\cite{Yeager2012}.\\
$^{f}$R-matrix, intermediate coupling, level averaged \cite{Witthoeft2009}.\\
$^{g}$Screening Constant by Unit Nuclear Charge (SCUNC) approximation \cite{Sakho2011,Sakho2012,Sakho2013}. \\
\end{table}

\begin{figure}
\begin{center}
\includegraphics[scale=2.5,height=14.0cm,width=16.0cm]{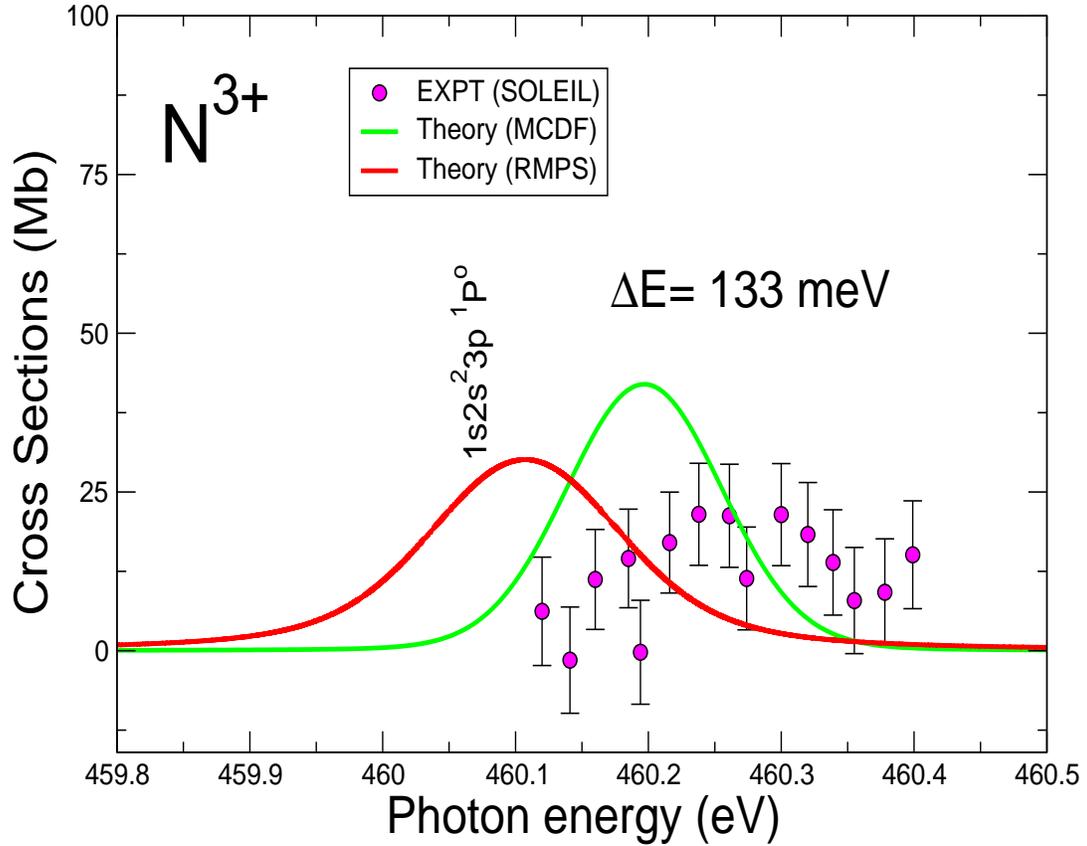}
\caption{\label{fig:56meV-1s-3p}(Colour online) Photoionization cross sections for Be-like atomic nitrogen (N$^{3+}$) ions measured 
								with a 133 meV band pass at the SOLEIL radiation facility for the 1s $\rightarrow$ 3p resonance. 
								Solid circles :  total photoionization. 
								The error bars give the statistical uncertainty of the experimental data. 
								The MCDF (solid green line) and R-matrix (solid red line) calculations shown are convolution with a Gaussian
								 profile of 133 meV FWHM and an appropriate weighting of the ground and metastable states 
								 (see text for details) to simulate the measurements.}
\end{center}
\end{figure}

\subsection{R-matrix: Li-like nitrogen}
For this Li-like system intermediate-coupling photoionization cross section 
calculations were performed using the semi-relativistic Breit-Pauli approximation 
which allows for relativistic effects to be included in a similar manner 
to our previous work on Li-like, boron \cite{Mueller2010} and carbon \cite{Mueller2009} ions.
As in our previous work, radiation-damping \cite{damp} effects were 
also included within the confines of the R-matrix approach \cite{rmat,codes} 
for completeness \cite{rmat,damp,dubau2013}. 
An appropriate number of N$\rm ^{5+}$ residual ion states 
(19 $LS$, 31 $LSJ$ levels) were included in our intermediate coupling calculations. 
 The n=4 basis set of  N$\rm ^{5+}$ orbitals obtained  
 from the  atomic-structure code CIV3 \cite{Hibbert1975} were used to represent the wavefunctions.  
Photoionization cross-section calculations were then performed in intermediate coupling for the
$\rm 1s^22s~^2S_{1/2}$ initial state of the N$\rm ^{4+}$ ion in order to incorporate
relativistic effects via the semi-relativistic Breit-Pauli approximation.

For cross section calculations He-like $LS$ states were retained:
$\rm 1s^2~^1S$,  $\rm 1sns~^{1,3}S$, 
$\rm 1snp~^{1,3}P^{\,\circ}$, $\rm 1snd~^{1,3}D$,  
and $\rm 1snf~^{1,3}F^{\,\circ}$, n $\leq$ 4,
of the N$\rm ^{5+}$ ion core giving rise to 31 $LSJ$ states 
in the intermediate  close-coupling expansions for
the $J$=1/2 initial scattering symmetry of the Li-like  N$^{4+}$ ion. 
The n=4 pseudo states are included in an attempt to account for core relaxation,  
electron correlations effects and the infinite number of states (bound and continuum) 
left out by the truncation of the close-coupling expansion in our work.
For the structure calculations of the residual
N$\rm ^{5+}$ ion, all n=3 physical orbitals  and n=4 correlation orbitals were included in the 
multi-configuration-interaction target wavefunctions expansions used to describe the states. 

The Hartree-Fock $\rm 1s$ and $\rm 2s$ orbitals of Clementi and Roetti
\cite{Clementi1974} together with the n=3 
 orbitals were determined by energy optimization on the appropriate
spectroscopic state using the atomic structure code CIV3
\cite{Hibbert1975}.  The n=4 correlation (pseudo) orbitals were determined by energy 
optimization on the ground state of this ion. 
All the states of the N$\rm^{5+}$ ion were then represented
by using multi-configuration interaction wave functions.  The Breit-Pauli
$R$-matrix approach was used to calculate the energies 
of the N${\rm ^{4+}}(LSJ)$ bound states and the subsequent photoionization cross sections. 
A minor shift ($<$  0.1 \%) of the theoretical energies for the N$\rm ^{5+}$ ion core states to experimental values 
\cite{NIST2012} was made so that they would be in agreement with available 
experimental thresholds.  Both double and triple promotion models 
for the scattering wave functions were investigated.
The accuracy of the bound initial state wavefunction is more
difficult to assess. Earlier $LS$ coupling calculations 
of Charro and co-workers \cite{Charro2000} gave a
value of 7.1909 Rydbergs, for the $\rm1s^22s~ ^2S_{1/2}$ bound state.
From our R-matrix calculation we obtain values  of 7.19301 Rydbergs (double electron promotions), 
and 7.19332 Rydbergs (triple electron promotions) compared to  
a value of 7.19479 Rydbergs from the NIST tabulation \cite{NIST2012}.
A discrepancy of 24 meV and 20 meV respectively with experiment.
Photoionization cross sections out of the Li-like nitrogen ion 
N$\rm ^{4+}$ (1s$\rm ^2$2s $\rm ^2$S$_{\rm 1/2}$ )
ground-state  were then obtained for total angular momentum scattering symmetries of 
$J$ = 1/2 and $J$= 3/2, odd parity, that contribute to the total.

The $R$-matrix method \cite{rmat,codes,damp} was used to determine 
all the photoionization cross sections for the initial ground state in $LS$ and intermediate-coupling. 
In previous R-matrix work on Li-like boron \cite{Mueller2010} and Li-like carbon \cite{Mueller2009} ions, 
two and three-electron promotions scattering models were both
employed yielding similar results.  A similar tendency is found here when
compared with the high resolution SOLEIL experimental measurements. We illustrate only
the R-matrix results from the triple-promotion scattering model with the SOLEIL experimental measurements.

The electron-ion collision work was carried out with twenty
continuum functions and a boundary radius of 6.8 Bohr radii. 
For the $\rm ^2S_{1/2}$ initial state the outer region electron-ion collision 
problem was solved using an appropriate fine 
energy mesh of  2.0$\times$ 10$^{-7}$ Rydbergs ($\approx$  2.72 $\mu$eV) 
to delineate the resonance features in the cross sections. 

For Li-like atomic nitrogen, the QB technique (applicable to atomic and molecular complexes) 
of Berrington and co-workers \cite{keith1996,keith1998,keith1999} 
was used to determine the resonance parameters and
averaging was performed over final total angular
momentum $J$ values. All the resonance parameters 
for the Li-like atomic nitrogen ion are presented in Table \ref{reson2}.  
Finally, in order to compare directly with experiment, 
the theoretical cross section was convoluted with a Gaussian 
function of appropriate full-width half-maximum (125 meV FWHM) to simulate 
the energy resolution of the measurements. The experimental and theoretical 
results for this Li-like atomic nitrogen ion are presented in figure 5.

\begin{figure}
\begin{center}
\includegraphics[scale=1.5,height=14.0cm,width=16.0cm]{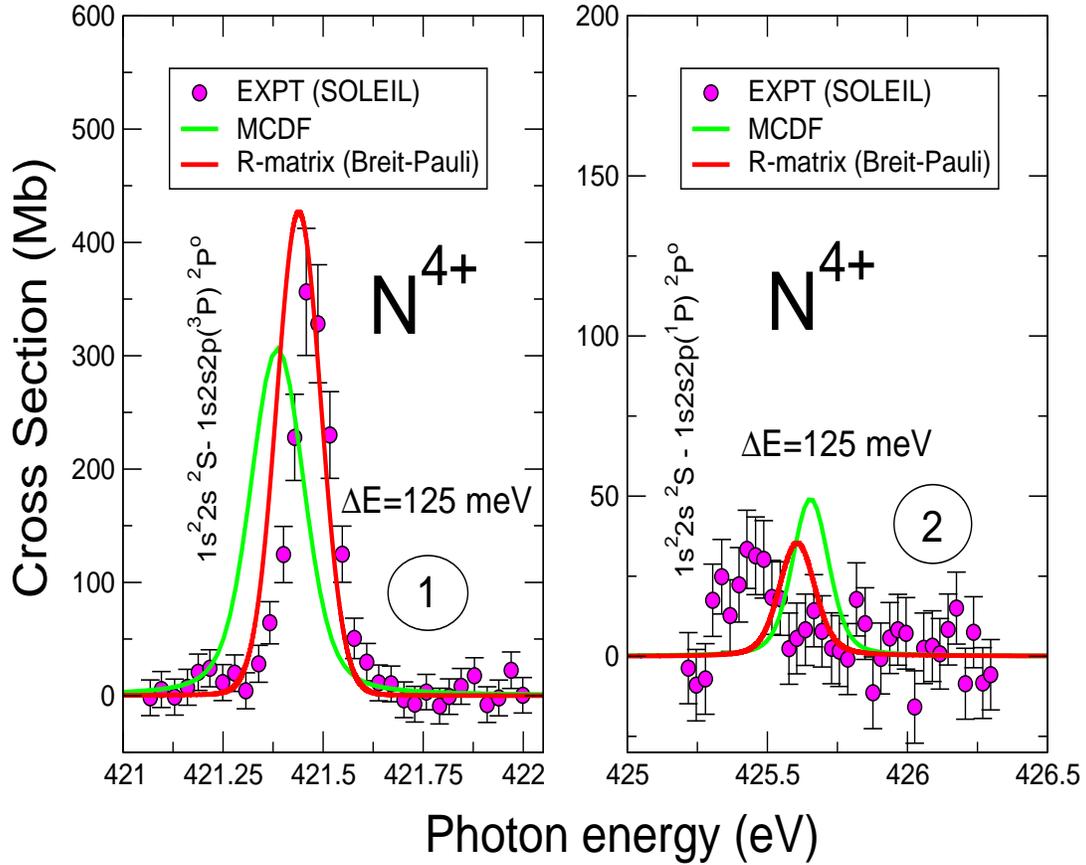}
\caption{\label{fig2:120meV}(Colour online) Photoionization cross sections for Li-like atomic nitrogen (N$^{4+}$) ions measured 
								with a 125 meV FWHM band pass at the SOLEIL radiation facility. 
								Solid circles : absolute total photoionization cross sections. 
								The error bars give the  statistical uncertainty of the experimental data. 
								 R-matrix  (solid red line, 31 levels) 
								 intermediate coupling, MCDF (solid green line),
								calculations shown are convolution with a Gaussian
								 profile of 125 meV FWHM to simulate the measurements. 								 
								 Table \ref{reson2} gives the resonances~\mycirc{1}  - \mycirc{2} and their parameters.}
\end{center}
\end{figure}

\section{Results and Discussion}\label{sec:Results}
Figures 1 to 4 compare our experimental cross sections (solid circles) with theoretical predictions made from the
MCDF (green curve) and R-matrix (red curve) methods for Be-like atomic nitrogen (N$^{3+}$).  Similarly in figure 5 
we present those for Li-like atomic nitrogen (N$^{4+}$) ions.  For the Be-like atomic nitrogen (N$^{3+}$) ion, 
the experimental cross section has been recorded with different band widths, ranging from 38 to 133 meV, 
presented in figures 1 to 4 respectively. Each cross section is obtained from the weighted mean 
of several sweeps (from 3 to 8 following increasing resolution). The assignment, resonance excitation energy,  Auger
width and strengths of all the n=2 observed lines are summarised in Tables \ref{reson} and \ref{reson2}. 
For the determination of the experimental widths, each individual sweep has been fitted separately by  Voigt profiles 
to avoid any possible shift in the energy delivered by the monochromator. Then the final width of the 
Lorentzian component was obtained from the weighted mean of the individual Lorentzian width 
determined from each fit. For the N$^{3+}$ ion, splitting of the $J$ components of the initial state \cite{NIST2012} 
was also taken into account, assuming a statistical distribution of the levels. The experimental and 
theoretical oscillator strengths were obtained from the area under the lines.

For Be-like atomic nitrogen ions, three peaks were observed in the experimental spectrum in the 
photon energy range 412 eV -- 415 eV investigated.
Figure 1, shows the experimental and theoretical results obtained at a resolution of 111 meV,
 figure 2 illustrates the results at the higher resolution of 56 meV. The double peak around 412 eV is more visible 
 at the higher resolution of 38 meV, see figure 3.  Due to the presence of metastable states in the beam of ions 
 for N$^{3+}$, in order to make a true comparison of theory with experiment, the theoretical cross sections were convoluted 
 using a Gaussian profile function of the appropriate width and an admixture of  40 \% metastable 
 and 60 \% ground state ions used  to simulate experiment. From our results illustrated in figures 1 -- 4
 it is seen over the entire photon energy region investigated, excellent agreement is achieved with the present 
 R-matrix calculation, both on the absolute cross sections scale and for resonance energies
 positions. The MCDF calculations show less favourable agreement, as the double peak resonances 
 due to the $\rm 1s^22s2p ~ ^3P^{\circ}$  metastable state lies over 1.46 eV below the experimental locations. 
 In figures 1 -- 3,  for Be-like nitrogen, we have shifted the 
 MCDF cross section calculations for the $\rm 1s^22s2p ~ ^3P^{\circ}$ 
 metastable  up by 1.46 eV in order to match the experimental measurements.
Table \ref{reson}, presents the resonance parameters for all three resonances 
found in the Be-like atomic nitrogen experimental measurements. We note that our present 
RMPS estimates for the position of all three resonances are in excellent agreement with experiment. 

 Estimates for the resonance energies, of Be-like nitrogen made using the 
screening constant by unit nuclear charge (SCUNC) empirical fitting approach \cite{Sakho2011} 
show satisfactory agreement with the more sophisticated theoretical methods and experiment.  
In the case of the Auger widths, apart from the first  $\rm 1s2s2p^2[^4P] ^3P$ resonance we see they are all in respectable 
agreement with experiment.  We note that the more sophisticated theoretical methods such as R-matrix and the Saddle 
Point Method (SPM) consistently give values in agreement with each other and in general with experiment.  
We point out that  in recent {\it K}-shell measurements  for Be-like boron (e.g., B$^{+}$ \cite{Mueller2013}) 
the Auger width for the $\rm 1s2s2p^2[^4P] ^3P$ gave a value of $\sim$ 10 meV, consistent with theory. 
So we suspect that even higher resolution than the present work performed at SOLEIL  
would be required to fully resolve the double peak structure in the N$^{3+}$ spectrum at about 412 eV.

Figure 4 shows the SOLEIL data  taken in the region of the $\rm 1s^22s^2~^1S \rightarrow 1s2s^23p~^1P^{\circ}$ 
resonance located around 460 eV at an energy resolution of 133 meV.  Due to the limited experimental data taken in this region (as illustrated 
in figure 4), it was not possible to fit the experimental data to extract a reliable Auger width.  We find an experimental value for the energy
of this resonance to be 460.280 $\pm$ 0.04 eV, a strength of 6.30 $\pm$ 2.6 which compares favourably with
theoretical predictions from the R-matrix with pseudo-states method (RMPS) value of 460.107 eV with 
a resonance strength of 6.59. Furthermore, the R-matrix with pseudo-states method (RMPS)  
provides an Auger width of 72 meV for this 1s $\rightarrow$ 3p resonance.
For the $\rm 1s \rightarrow 3p$ resonance a position of 460.019 $\pm$ 0.045 eV 
with an Auger width of 87 meV was determined from the empirical fitting SCUNC approximation \cite{Sakho2011}.
 The optical potential R-matrix method \cite{Witthoeft2009}) for this same resonance gave 
 a value of 462.373 eV with an Auger width of 68 meV.  The present MCDF work gave
values of 460.189 eV for the position and 80 meV for the Auger width.
No other values appear to be available in the literature.  We note that numerical values 
of cross sections from previous R-matrix work  \cite{Witthoeft2009} are
available for only the ground-state for this Be-like ion and as such 
similar comparisons cannot be made in figures 1-4. 
%
%
%
%
%
\begin{table}
\caption{\label{reson2} 	Li-like nitrogen (N$^{4+}$),  present experimental and 
					theoretical results for the resonance energies $E_{\rm ph}^{\rm (res)}$ (eV),
            				natural line-widths $\Gamma$ (meV) and resonance strengths $\overline{\sigma}^{\rm PI}$ (in Mb eV),
           				for the  $\rm 1s^22s\, ^2$S\,$\rightarrow$\, $\rm 1s[2s2p~ ^{1,3}P]\, ^2$P$^{\circ}$  core excited 
           				states of the N$^{4+}$ ion, in the photon energy region 
           				420 eV to  426 eV compared with previous investigations. The intermediate-coupling 
           				results have been averaged over the fine structure levels 
           				to compare with experiment and with other theoretical methods.  
           				The  error in the calibrated photon energy is estimated to be $\pm$ 30 meV
				          for the resonance energies and the experimental resolution was 125 meV determined
				          from multi-function Voigt fits to the measurements.} 
 \lineup
  \begin{tabular}{ccr@{\,}c@{\,}llcl}
\br
 Resonance    & & \multicolumn{3}{c}{SOLEIL/others}                                       & \multicolumn{1}{c}{R-matrix} 		& \multicolumn{2}{c}{MCDF/others}\\
 (Label)            & & \multicolumn{3}{c}{(Experiment$^{\dagger}$)}      & \multicolumn{1}{c}{(Theory)} 		& \multicolumn{2}{c}{(Theory)}\\
 \ns
 \mr
 $\rm 1s^22s\, ^2$S\,$\rightarrow$\,  $\rm 1s[2s2p~ ^3P]\, ^2$P$^{\circ}$ 			
   				& $E_{\rm ph}^{\rm (res)}$     &   		& 421.472$\pm$ 0.03$^{\dagger}$   & 				&421.448$^{a}$    	 &	&   421.390$^{b}$  \\
  ~\mycirc{1}		&          	 		               &   		& 421.521$\pm$ 0.05$^{\ddagger}$ &				&420.612$^{f}$        	 &	&   420.940$^{c}$  \\
				&          	 		               &   		& 421.228$\pm$ 0.05$^{\ast}$		&				&     				 &	&   421.321$^{d}$  \\ 
				&          	 		               &   		& 421.120	$\pm$ 0.07$^{\S}$	 	&			    	&				 &	&   421.572$^{e}$  \\ 
				&          	 		               &   		& 							&				&     				 &	&   421.169$^{g}$ \\			   
				&          	 		               &   		& 							&				&     				 &	&   421.390$^{h}$ \\			   
				&          	 		               &   		& 							&				&     				 &	&   421.482$^{i}$ \\			   
				\\ 
                                     & $\Gamma$                             & \;\0\0\         &   11 $\pm$ 8			            	&     				&  \0\0\04$^{a}$  	 &	&       \0\05$^{c}$  \\
		 	         &                                                  &   		&  						         &				&   \0\0\06$^{f}$    	 &	&        \0\04$^{d}$ \\
		 	         &                                                  &   		&  						         &				&     		         		 &	&        \0\06$^{g}$ \\
		 	         &                                                  &   		&  						         &				&     		         		 &	&        \0\06$^{h}$\\
		 	         &                                                  &   		&  						         &				&     		         		 &	&       \0\05$^{i}$ \\
			         \\
				&$\overline{\sigma}^{\rm PI}$&			&  47.3 $\pm$ 7.4		        		&				& \0\0 60$^a$		 &	& 	\\
 	   		       \\
  $\rm 1s^22s\, ^2$S\,$\rightarrow$\, $\rm 1s[2s2p~ ^1P]\, ^2$P$^{\circ}$		
                                     & $E_{\rm ph}^{\rm (res)}$     &	 		 &425.449 $\pm$ 0.03$^{\dagger}$  & 			&425.606$^{a}$    	 &	&    425.654$^{b}$ \\
  ~ \mycirc{2}	         &           				     &		   	 &425.624 $\pm$ 0.40$^{\ast}$		&				&424.823$^{f}$    	 &	&    426.020$^{c}$ \\
  			         &           				     &		   	 &424.890  $\pm$ 0.15$^{\S}$		&				&     		         		 &	&    425.421$^{d}$ \\
  			         &           				     &		   	 &  						  	&				&     		         		 &	&    425.770$^{e}$ \\
 			         &           				     &		   	 &  						  	&				&     		         		 &	&    425.530$^{h}$ \\
			         &           				     &		   	 &  						  	&				&     		         		 &	&    425.666$^{i}$  \\
			         \\
			          & $\Gamma$ 			     & \;\0\0\    	 & --				     			&				& \0\042$^{a}$  	 &	&         \016$^{c}$ \\
 	    			&           				     &   		 &  						         &				& \0\053$^{f}$    	 &	&	  \042$^{d}$\\ 
  	    			&           				     &   		 &  						         &				&     		 		 &	&         \043$^{h}$ \\ 
	    			&           				     &   		 &  						         &				&     		 		 &	&         \044$^{i}$\\ 
				\\  
				&$\overline{\sigma}^{\rm PI}$&			 & 7.3 $\pm$ 2.5				&				& \0\06.5$^{a}$	 	&	&	 \\
	   		       \\
 \br
\end{tabular}
~\\
$^{\dagger}$SOLEIL, present measurements.\\
$^{\ddagger}$EBIT,  measurements \cite{EBIT1999}.\\
$^{\ast}$Laser produced plasmas (LPP), measurements \cite{Tondello1977}.\\
$^{\S}$Electron spectroscopy in ionÐatom collisions, measurements \cite{Niehaus1987}.\\
$^{a}$R-matrix, intermediate-coupling (31 levels) present results, level averaged. \\
$^{b}$Multi-configuration Dirac-Fock (MCDF) method, present work, $^{c}$MCDF \cite{Chen1983,Chen1986}.\\
$^{d}$Saddle-point method with complex-rotation (SPM-CR) \cite{Davis1989}.\\
$^{e}$Intermediate-coupling, semi-relativistic method \cite{Gabriel1972}.\\
$^{f}$R-matrix, intermediate coupling, results, level averaged \cite{Witthoeft2009}.\\
$^{g}$Complex scaled multi-reference configuration interaction method (CMR-CI) \cite{Zhang2013}.\\
$^{h}$Saddle-point method with R-matrix (SPM-RM). \cite{Wu1991}.\\
$^{i}$Screening Constant by Unit Nuclear Charge (SCUNC) approximation \cite{Sakho2011,Sakho2012,Sakho2013}. \\
\end{table}

Table \ref{reson2} presents the resonance parameters for the two resonances 
observed in the Li-like atomic nitrogen experimental measurements illustrated in figure 5.
We note that only the first narrow peak in the Li-like experimental spectrum located 
at 421.472 $\pm$ 0.03 eV was able to be properly fitted, giving a Auger line width of 11 $\pm$ 8 meV.
Both the sophisticated  R-matrix and Saddle point methods 
are in agreement with each other and the experimental value 
for the location of this resonance.  Theoretical predictions from R-matrix calculations 
and other methods for the Auger width all lie within the experimental error estimate for this resonance.
The second peak in the  Li-like atomic nitrogen photoionization cross section, located at 425.449 $\pm$ 0.03 eV, 
had limited experimental data taken on the lower energy side of the peak and  due to the rather large noise level 
(large scatter in the experimental cross section data)  made it impossible to fit.  
The noise level was simply just too large to do a proper fit of this peak. 
For the natural line width  of this $\rm 1s[2s2p~ ^1P]\, ^2$P$^{\circ}$ resonance, 
R-matrix calculations give a value of 42 meV, present MCDF  estimates of 45 meV, 
and the Saddle point method 42 meV. 
From previous R-matrix work  \cite{Witthoeft2009} we have extracted resonance 
positions and widths.  For the first resonance we find an energy of 420.612 eV
and for the second resonance an energy of 424.823 eV,  a discrepancy of 
about 0.9 eV and  0.6 eV lower than the present experiment, 
with values of 6 meV and 53 meV for the respective widths.

It is also see that the present R-matrix calculations yield a resonance strength of 6.5 Mb eV, 
MCDF estimates a value of 9 Mb eV and both theoretical values 
are in suitable agreement with the experimental measurement of 7.3 $\pm$ 2.5.
For the positions of both resonances, the  
theoretical values from the RMPS calculations are in more favourable agreement 
with the SOLEIL synchrotron measurements than 
either the earlier EBIT   \cite{EBIT1999}, the Laser 
produced plasmas (LPP)  \cite{Tondello1977} or 
those determined from electron spectroscopy in ion - atom 
collisions \cite{Niehaus1987} measurements. 
Table \ref{reson2} indicates that estimates for the resonance energies 
and Auger widths of Li-like nitrogen made using the 
screening constant by unit nuclear charge (SCUNC) 
empirical fitting approach \cite{Sakho2011,Sakho2012,Sakho2013} 
are in satisfactory agreement with the more 
sophisticated theoretical methods and experiment.  

\section{Conclusions}\label{sec:Conclusions}
{\it K}-shell photoionization for Be-like and Li-like atomic nitrogen ions, 
respectively N$^{3+}$ and N$^{4+}$, has been performed
using state-of-the-art experimental and theoretical methods in the vicinity of their 
respective {\it K}-edges. To our knowledge this would appear to be the first time 
high-resolution spectroscopy has been performed  
(at a photon energy resolution of  respectively, 38 meV, 56 meV, 111 meV, 133 meV FWHM  
for Be-like and 125 meV FWHM for Li-like atomic nitrogen ions). 
The measurements at the SOLEIL synchrotron radiation facility, in Saint-Aubin, France, 
cover the photon energy ranges 410 eV -- 415 eV and 460 -- 460.4 eV for Be-like 
and 420 eV -- 426 eV for Li-like atomic nitrogen ions.  
The strong peaks found in the respective PI cross sections in the 
 photon energy regions studied are identified as the 1s $\rightarrow$ 2p and  1s $\rightarrow$ 3p transitions  
 in the Be-like and 1s $\rightarrow$ 2p transitions in the Li-like  {\it  K}-shell spectrum that are assigned spectroscopically.
 All the n=2 resonance parameters have been tabulated in Table \ref{reson} for Be-like 
 and in  Table \ref{reson2} for Li-like atomic nitrogen ions.
 For the observed peaks, respectable agreement is seen with the present theoretical and
experimental results both on the photon-energy scale and the absolute cross section scale.
Some differences are highlighted and discussed.
 
The strength of the present study is the high resolution of  the spectra along with 
 theoretical predictions made using state-of-the-art MCDF and R-matrix  methods.  
 The present results have been compared with high resolution 
 experimental measurements made at the SOLEIL 
 synchrotron radiation facility and with other theoretical methods so
 would be suitable to be incorporated into astrophysical modelling codes like 
 CLOUDY \cite{Ferland1998,Ferland2003}, XSTAR \cite{Kallman2001} 
 and AtomDB \cite{Foster2012} used to numerically
simulate the thermal and ionization structure of ionized astrophysical nebulae. 

\ack
The experimental measurements were performed on the 
PLEIADES beamline \cite{Pleiades2010,Miron2013}, at the SOLEIL Synchrotron 
radiation facility in Saint-Aubin, France.  The authors would like to thank the SOLEIL staff and, 
in particular those of the PLEIADES beam line for their helpful assistance.
B M McLaughlin acknowledges support from the US National Science Foundation through a grant to ITAMP
at the Harvard-Smithsonian Center for Astrophysics, the RTRA network {\it Triangle de le Physique} 
and a visiting research fellowship from Queen's University Belfast. 
I Sakho acknowledges the hospitality of the Universit\'{e} Paris-Sud and support 
from the University Assane Seck of Ziguinchor during a recent visit.
We thank Dr John C Raymond and Dr Randall K Smith at the Harvard-Smithsonian Center 
for Astrophysics for discussions on the astrophysical applications and Dr Javier Garcia 
for numerical values of the cross sections  \cite{Witthoeft2009}.
We gratefully acknowledge Dr. Wayne C Stolte for assistance in fitting  and 
interpreting the SOLEIL experimental measurements with the WinXAS software.
The computational work was carried out at the National Energy Research Scientific
Computing Center in Oakland, CA, USA, the  Kraken XT5 facility at the National Institute 
for Computational Science (NICS) in Knoxville, TN, USA
and at the High Performance Computing Center Stuttgart (HLRS) of the University of Stuttgart. 
We thank Stefan Andersson from Cray Research for his advice and assistance with 
the implementation of the parallel R-matrix codes on the Cray-XE6 at HLRS.
The Kraken XT5 facility is a resource of the Extreme Science and Engineering Discovery Environment (XSEDE), 
which is supported by National Science Foundation grant number OCI-1053575.
%
%
%
%

\bibliographystyle{iopart-num}
\bibliography{nions}

\providecommand{\newblock}{}
\begin{thebibliography}{100}
\expandafter\ifx\csname url\endcsname\relax
  \def\url#1{{\tt #1}}\fi
\expandafter\ifx\csname urlprefix\endcsname\relax\def\urlprefix{URL }\fi
\providecommand{\eprint}[2][]{\url{#2}}

\bibitem{McLaughlin2001}
{McLaughlin B M} 2001 {} {\em {Spectroscopic Challenges of Photoionized
  Plasma}\/} ({\em ASP Con$f$. Series\/} vol \textbf{247}) ed {Ferland, G and
  Savin D W} (San Francisco, CA: Astronomical Society of the Pacific) p~87

\bibitem{Brickhouse2010}
{Foster A, Smith R, Brickhouse N, Kallman T, Witthoeft M} 2010 {\em {Space Sci.
  Rev.}\/} {\bf \textbf{157}} 135

\bibitem{Kallman2010}
{Kallman T~R} 2010 {\em {Space Sci. Rev.}\/} {\bf \textbf{157}} 177

\bibitem{Quinet2011}
{Quinet P, Palmeri P, Mendoza C, Bautista M, Garcia J, Witthoeft M, Kallman
  T~R} 2011 {\em {J Elect. Spec. and Relat. Phenom.}\/} {\bf \textbf{184}} 170

\bibitem{McLaughlin2013}
{McLaughlin B M and Ballance C P} 2013 {Photoionization, Fluorescence and
  Inner-shell Processes} {\em {McGraw-Hill Yearbook of Science and Technology
  2013}\/} ed {McGraw-Hill} (New York, USA: McGraw-Hill Inc) p 281

\bibitem{McLaughlin2010}
{Hasoglu M~F, Abdel-Naby Sh~ A, and Gorczyca T~W, Drake J~J and McLaughlin B~M}
  2010 {\em {Astrophys. J.}\/} {\bf \textbf{724}} 1296

\bibitem{Skinner2010}
{Skinner S~L, Zhekov S~A,G\"udel M, Schmutz W, Sokal K~R} 2010 {\em {Astronom.
  J.}\/} {\bf \textbf{139}} 825

\bibitem{Garcia2011}
{Garcia J, Ram\'irez, J~M, Kallman T~R, Witthoeft M, Bautista M~A, Mendoza C,
  Palmeri P and Quinet P} 2011 {\em {Astrophys. J}\/} {\bf \textbf{731}} L15

\bibitem{Witthoeft2009}
{Garcia J, Kallman T~R, Witthoeft M, Behar E, Mendoza C, Palmeri P, Quintet P,
  Bautista M and Klapisch M} 2009 {\em {Astrophys. J Supp. Ser.}\/} {\bf
  \textbf{185}} 477

\bibitem{McLaughlin2011}
{Sant'Anna M~M, Schlachter A~S, \"{O}hrwall G, Stolte W~C, D W Lindle D~W and
  McLaughlin B~M } 2011 {\em {Phys. Rev. Letts.}\/} {\bf \textbf{107}} {\it
  033001}

\bibitem{Soleil2011}
{Gharaibeh M F, Bizau J M Cubaynes D, Guilbaud S, Hassan N El, Shorman M M Al,
  Miron C, Nicolas C, Robert E, Blancard C and McLaughlin B M} 2011 {\em {J.
  Phys. B: At. Mol. Opt. Phys.}\/} {\bf \textbf{44}} 175208

\bibitem{Diaz2007}
{ P\'{e}rez-Montero E and D\'{õ}az A} 2007 {\em {Mon. Not. Roy. Astron.
  Soc.}\/} {\bf \textbf{377}} 1195

\bibitem{Bohigas2008}
{Bohigas J} 2005 {\em {Astrophys. J}\/} {\bf \textbf 674} 954

\bibitem{Verner2005}
{Verner E, Bruhweiler F and Gull T} 2005 {\em {Astrophys. J}\/} {\bf
  \textbf{624}} 973

\bibitem{Sternberg2004}
{Gnat O and Sternberg A} 2004 {\em {Astrophys. J.}\/} {\bf \textbf{608}} 229

\bibitem{Ferland1998}
{Ferland G J, Korista K T, Verner D A, Ferguson J W, Kingdon J B and Verner E
  M} 1998 {\em {Pub. Astron. Soc. Pac.(PASP)}\/} {\bf \textbf{110}} 761

\bibitem{Ferland2003}
{Ferland G J} 2003 {\em {Ann. Rev. of Astron. \& Astrophys.}\/} {\bf
  \textbf{41}} 517

\bibitem{Mewe2001}
{Mewe R {\it et al}} 2001 {\em {Astron. \& Astrophys.}\/} {\bf \textbf 368} 888

\bibitem{Ness2001}
{Ness J-U {\it et al}} 2001 {\em {Astron. \& Astrophys.}\/} {\bf \textbf 367}
  282

\bibitem{stelzer2002}
{Stelzer B, {\it et al}} 2002 {\em {Astron. \& Astrophys.}\/} {\bf
  \textbf{392}} 585

\bibitem{Ness2007}
{Ness J-U {\it et al}} 2007 {\em {Astrophys. J}\/} {\bf \textbf 665} 1334

\bibitem{Ness2011}
{Ness J-U {\it et al}} 2011 {\em {Astrophys. J}\/} {\bf \textbf 733} 70

\bibitem{Kaastra2002}
{Kaastra J {\it et al}} 2002 {\em {Astron. \& Astrophys.}\/} {\bf \textbf{386}}
  427

\bibitem{EBIT1999}
{Beiersdorfer P {\it et al}} 1999 {\em {Rev. Sci. Instrum.}\/} {\bf \textbf 70}
  276

\bibitem{Rice1997}
{Rice J E {\it et al}} 1997 {\em {Phys. Plasmas}\/} {\bf \textbf 4} 1605

\bibitem{bizau2005}
{Bizau J M {\it et al}} 2005 {\em {Astron. \& Astrophys.}\/} {\bf \textbf{439}}
  387

\bibitem{Scully2006}
{Scully S W J, {\'A}lvarez I, Cisneros C, Emmons E D, Gharaibeh M~F, Leitner D,
  Lubell M S, M\"{u}ller A, Phaneuf R A, P\"{u}ttner R, Schlachter A S,
  Schippers S, Ballance C P and McLaughlin B M} 2006 {\em {J. Phys. B: At. Mol.
  Opt. Phys.}\/} {\bf 39} 3957

\bibitem{Scully2007}
{Scully S W J, {\'A}lvarez I, Cisneros C, Emmons E D, Gharaibeh M~F, Leitner D,
  Lubell M S, M\"{u}ller A, Phaneuf R A, P\"{u}ttner R, Schlachter A S,
  Schippers S, Ballance C P and McLaughlin B M} 2007 {\em {J. Phys. Conf. Ser.
  }\/} {\bf \textbf{58}} 387

\bibitem{DPI2013}
{McLaughlin B M} 2013 {\em {J. Phys. B: At. Mol. Opt. Phys.}\/} {\bf
  \textbf{46}} 075204

\bibitem{Diehl1996}
{Diehl S {\it et al}} 1996 {\em {Phys. Rev. Letts}\/} {\bf \textbf{76}} 3915

\bibitem{Mueller2010}
{M{\"u}ller A, Schippers S, Phaneuf R~A, Scully S W J, Aguilar A, Cisneros C,
  Gharaibeh M~F, Schlachter A~S and McLaughlin B~M} 2010 {\em {J. Phys. B: At.
  Mol. Opt. Phys.}\/} {\bf 43} 135602

\bibitem{Mueller2009}
{M{\"u}ller A, Schippers S, Phaneuf R~A, Scully S W J, Aguilar A, Covington A
  M, {\'A}lvarez I, Cisneros C, Emmons E D, Gharaibeh M~F, Schlachter A~S,
  Hinojosa G and McLaughlin B~M} 2009 {\em {J. Phys. B: At. Mol. Opt. Phys.}\/}
  {\bf 42} 235602

\bibitem{Mueller2013}
{M{\"u}ller A, Schippers S, Phaneuf R~A, Scully S W J, Aguilar A, Cisneros C,
  Gharaibeh M~F, Schlachter A~S and McLaughlin B~M} 2013 {\em {J. Phys. B: At.
  Mol. Opt. Phys.}\/} {\bf ~} {\it in preparation}

\bibitem{Scully2005}
{Scully S W J, Aguilar A, Emmons E D, Phaneuf R A, Halka M, Leitner D, Levin J
  C, Lubell M S, P\"{u}ttner R, Schlachter A S, Covington A M, Schippers S,
  M\"{u}ller A and McLaughlin B M} 2005 {\em {J. Phys. B: At. Mol. Opt.
  Phys.}\/} {\bf \textbf{38}} 1967

\bibitem{Schlachter2004}
{Schlachter A S, Sant'Anna M M, Covington A M, Aguilar A, Gharaibeh M~F, Emmons
  E D, Scully S W J, Phaneuf R A, Hinojosa G, {\'A}lvarez I, Cisneros C,
  M\"{u}ller A and McLaughlin B M} 2004 {\em {J. Phys. B: At. Mol. Opt.
  Phys.}\/} {\bf \textbf{37}} L103

\bibitem{Kawatsura2002}
{Kawatsura K {\it et al}} 2002 {\em {J. Phys. B: At. Mol. Opt. Phys. }\/} {\bf
  \textbf{35}} {\it 4147}

\bibitem{Yamaoka2001}
{Yamaoka H {\it et al}} 2001 {\em {Phys. Rev.}\/} {\bf \textbf{A~65}} {\it
  012709}

\bibitem{Krause1994}
{Krause M O} 1994 {\em {Nucl. Instr. and Meth. in Phys. Res. B}\/} {\bf
  \textbf{87}} 178

\bibitem{Menzel1996}
{Menzel A, Benzaid S, Krause M, Caldwell C D, Hergenhahn U and Bissen M} 1996
  {\em {Phys. Rev. A}\/} {\bf \textbf{54}} R991

\bibitem{Stolte1997}
{Stolte W C, Samson J A R, Hemmers O, Hansen D, Whitfield S B and Lindle D W }
  1997 {\em {J. Phys. B: At. Mol. Opt. Phys.}\/} {\bf \textbf{30}} 4489

\bibitem{Stolte2013}
{McLaughlin B M, Ballance C P, Bown K P, Gardenghi D J and Stolte W C} 2013
  {\em {Astrophys. J}\/} {\bf \textbf{771}} L8

\bibitem{Phaneuf2011}
{Covington A M, Aguilar A, Covington I R, Hinojosa G, Shirley, C A, Phaneuf
  R~A, {\'A}lvarez I, Cisneros C, Dominguez-Lopez I, Sant'Anna M M, Schlachter
  A~S, Ballance C P and McLaughlin B~M} 2011 {\em {Phys. Rev. A}\/} {\bf 84}
  013413

\bibitem{Simon2010}
{Simon M C {\it et al}} 2010 {\em {Phys. Rev. Letts.}\/} {\bf \textbf{105}}
  {\it 183001}

\bibitem{Ballance2012}
{McLaughlin B M and Ballance C P} 2012 {\em {J. Phys. B: At. Mol. Opt.
  Phys.}\/} {\bf \textbf{45}} 095202

\bibitem{Esteves2011}
{Esteves~D~A {\it et al}} 2011 {\em {Phys. Rev. A}\/} {\bf \textbf{84}} 013406

\bibitem{Sterling2011}
{Sterling N C {\it et al}} 2011 {\em {J. Phys. B: At. Mol. Opt. Phys.}\/} {\bf
  \textbf{44}} 025701

\bibitem{Bizau2011}
{Bizau J M, Blancard C, Coreno M, Cubaynes D, Dehon C, Hassan N E, Folkmann F,
  Gharaibeh M F, Giuliani A, Lemaire J, Milosavljevi A R, Nicolas C, and
  Thissen R} 2011 {\em {J. Phys. B: At. Mol. Opt. Phys.}\/} {\bf \textbf{44}}
  055205

\bibitem{McLaughlin2012}
{McLaughlin B M and Ballance C P} 2012 {\em {J. Phys. B: At. Mol. Opt.
  Phys.}\/} {\bf \textbf{45}} 085701

\bibitem{Hino2012}
{Hinojosa G, Covington A M, Alna'Washi G A, Lu M, Phaneuf R~A, Sant'Anna M M,
  Cisneros C, {\'A}lvarez I, Aguilar A, Kilcoyne, Schlachter A~S, Ballance C P
  and McLaughlin B~M} 2012 {\em {Phys. Rev. A}\/} {\bf 86} 063402

\bibitem{Soleil2013}
{Gharaibeh M F {\it et al}} 2013 {\em {J. Phys. B: At. Mol. \& Phys.}\/} {\bf
  \textbf{ }} in preparation for publication

\bibitem{rmat}
{Burke P G and Berrington K A} 1993 {\em {Atomic and Molecular Processes: An
  R-matrix Approach}\/} (Bristol, UK: IOP Publishing)

\bibitem{codes}
{Berrington K A, Eissner W and Norrington P~H} 1995 {\em {Comput. Phys.
  Commun.}\/} {\bf \textbf{92}} 290
  \urlprefix\url{http://amdpp.phys.strath.ac.uk/APAP}

\bibitem{Tondello1977}
{Nicolosi P and Tondello G} 1977 {\em {J. Opt. Soc. Am.}\/} {\bf \textbf{67}}
  1033

\bibitem{Niehaus1987}
{Mack M and Niehaus A} 1987 {\em {Nucl. Instrum. \& Methods in Phys. Res. B}\/}
  {\bf \textbf{23}} 291

\bibitem{Mueller2007}
{M{\"u}ller A, Schippers S, Phaneuf R~A, Kilcoyne A L D, Br{\"a}uning H,
  Schlachter A~S, Lu M and McLaughlin B~M} 2007 {\em {J. Phys.: Conf. Ser.}\/}
  {\bf \textbf{58}} 383

\bibitem{Mueller2010b}
{M{\"u}ller A, Schippers S, Phaneuf , Kilcoyne A L D, Br{\"a}uning H,
  Schlachter A S, Lu M and McLaughlin B~M} 2010 {\em {J. Phys. B: At. Mol. Opt.
  Phys.}\/} {\bf 43} 225201

\bibitem{Petrini1997}
{Petrini~D and de Ara\'ujo~ F~X} 1997 {\em {Astron. \& Astrophys.}\/} {\bf
  \textbf{326}} 870

\bibitem{Slater1960}
{Slater J~C} 1960 {\em {Quantum Theory of Atomic Structure}\/} (New York, USA:
  McGraw-Hill)

\bibitem{HS1963}
{Herman F and Skillman S} 1963 {\em {Atomic Structure Calculations}\/}
  (Englewood Cliffs, NJ, USA: Prentice-Hall))

\bibitem{Reilman1979}
{Reilman R~F and Manson S~T} 1979 {\em {Astrophys. J. Suppl. Ser.}\/} {\bf
  \textbf{40}} 85

\bibitem{Band1979}
{Band I~M, Kharitonov Y~I and Trzhaskovskaya M~B} 1979 {\em {At. Data Nucl.
  Data Tables}\/} {\bf \textbf{23}} 443

\bibitem{Verner1993}
{Verner D~A {\it et al}} 1993 {\em {At. Data Nucl. Data Tables}\/} {\bf
  \textbf{55}} 233

\bibitem{Berrington1997}
{Berrington K, Quigley L, and Zhang H L} 1997 {\em {J. Phys. B: At. Mol. \&
  Phys.}\/} {\bf \textbf{30}} 5409

\bibitem{Petrini1981}
{Petrini~D} 1981 {\em {J. Phys. B: At. Mol.}\/} {\bf \textbf{14}} 3839

\bibitem{Petrini1998}
{Stoica S, Petrini~D and Bely-Dubau F} 1998 {\em {Astron. \& Astrophys.}\/}
  {\bf \textbf{334}} L26

\bibitem{Charro2000}
{Charro E, Bell K L, Martin I and Hibbert A} 2000 {\em {Mon. Not. Roy. Astron.
  Soc.}\/} {\bf \textbf{313}} 247

\bibitem{damp}
{Robicheaux F, Gorczyca T W, Griffin D C, Pindzola M S and Badnell N R} 1995
  {\em {Phys. Rev. A}\/} {\bf 52} 1319

\bibitem{Burke2011}
{Burke P G} 2011 {\em {R-Matrix Theory of Atomic Collisions: Application to
  Atomic, Molecular and Optical Processes}\/} (New York, USA: Springer)

\bibitem{Sakho2011}
{Sakho I} 2011 {\em {Rad. Phys. Chem.}\/} {\bf \textbf{80}} 1295

\bibitem{Chen1983}
{Chen M H, Crasemann B and Hans M} 1983 {\em {Phys. Rev. A}\/} {\bf
  \textbf{27}} 544

\bibitem{Chen1985}
{Chen M H} 1985 {\em {Phys. Rev. A}\/} {\bf \textbf{31}} 1449

\bibitem{Pleiades2010}
{Travnikova O et al} 2010 {\em {Phys. Rev. Letts.}\/} {\bf \textbf{105}} 233001

\bibitem{Miron2013}
{Miron C {\it et al} 2013}
  \urlprefix\url{http://www.synchrotron-soleil.fr/portal/page/portal/Recherche/LignesLumiere/PLEIADES}

\bibitem{Sodhi1984}
{Sodhi R~N~S and Brion C~E} 1977 {\em {J. Electron Spectrosc. Relat.
  Phenom.}\/} {\bf \textbf{34}} 363

\bibitem{Kato2007}
{Kato M {\it et al}} 2007 {\em {J. Electron Spectrosc. Relat. Phenom.}\/} {\bf
  \textbf{160}} 39

\bibitem{Miron2012}
{Miron C {\it et al}} 2012 {\em {Nature Phys.}\/} {\bf \textbf{8}} 135

\bibitem{Kimberg2013}
{Kimberg V {\it et al}} 2013 {\em {Phys. Rev. X}\/} {\bf \textbf{3}} 011017

\bibitem{Sakho2012}
{Sakho I} 2012 {\em {Phys. Rev. A}\/} {\bf \textbf{86}} 052511

\bibitem{Sakho2013}
{Sakho I {\it et a}l} 2013 {\em {At. Data. Nuc. Data Tables}\/} {\bf
  \textbf{99}} 447

\bibitem{Bruneau1984}
{Bruneau J} 1984 {\em {J. Phys. B: At. Mol. Phys.}\/} {\bf \textbf{17}} 3009

\bibitem{ballance06}
{Ballance C P and Griffin D C} 2006 {\em {J. Phys. B: At. Mol. Opt. Phys.}\/}
  {\bf \textbf{39}} 3617

\bibitem{Clementi1974}
{Clementi E and Roetti C} 1974 {\em {At. Data Nucl. Data Tables}\/} {\bf
  \textbf{14}} 177

\bibitem{Hibbert1975}
{Hibbert A} 1975 {\em {Comput. Phys. Commun.}\/} {\bf \textbf{9}} 141

\bibitem{NIST2012}
{Kramida A E, Ralchenko Y, Reader J, and NIST ASD Team,} {NIST Atomic Spectra
  Database (version 5),} National Institute of Standards and Technology,
  Gaithersburg, MD, USA
  \urlprefix\url{http://physics.nist.gov/PhysRefData/ASD/levels_form.html}

\bibitem{Fano1968}
{Fano U and Cooper J W} 1968 {\em {Rev. Mod. Phys.}\/} {\bf \textbf{40}} 441

\bibitem{Shore1967}
{Shore B W} 1967 {\em {Rev. Mod. Phys.}\/} {\bf \textbf{39}} 439

\bibitem{Shore1967b}
{Shore B W} 1967 {\em {J. Opt. Soc. Am}\/} {\bf \textbf{57}} 881

\bibitem{Shore1968}
{Shore B W} 1967 {\em {Phys. Rev.}\/} {\bf \textbf{171}} 43

\bibitem{Ederer1971}
{Ederer D L} 1971 {\em {Phys. Rev. A}\/} {\bf \textbf{4}} 2263

\bibitem{Ederer1976}
{Ederer D L} 1976 {\em {Phys. Rev. A}\/} {\bf \textbf{14}} 1936 (E)

\bibitem{Morgan2008}
{Wright J D {\it et al}} 2008 {\em {Phys. Rev. A}\/} {\bf \textbf{77}} 062512

\bibitem{keith1996}
{Quigley L and Berrington K~A} 1996 {\em {J. Phys. B: At. Mol. Phys.}\/} {\bf
  \textbf{29}} 4529

\bibitem{keith1998}
{Quigley L, Berrington K A and Pelan J} 1998 {\em {Comput. Phys. Commun.}\/}
  {\bf 114} 225

\bibitem{keith1999}
{Ballance C P, Berrington K A and McLaughlin B M} 1999 {\em {Phys. Rev. A}\/}
  {\bf \textbf{60}} R4217

\bibitem{Lin2001}
{Lin S -H, Hsue C - S and Chung K T} 2001 {\em {Phys. Rev. A}\/} {\bf
  \textbf{64}} 012709

\bibitem{Lin2002}
{Lin S -H, Hsue C - S and Chung K T} 2002 {\em {Phys. Rev. A}\/} {\bf
  \textbf{65}} 032706

\bibitem{Yeager2012}
{Zhang S B and Yeager D L} 2012 {\em {Phys. Rev. A}\/} {\bf \textbf{85}} 032515

\bibitem{dubau2013}
{Zabaydullin Z and Dubau J} 2013 {\em {J. Phys. B: At. Mol. \& Opt. Phys.}\/}
  {\bf \textbf{46}} 075005

\bibitem{Chen1986}
{Chen~M~H} 1986 {\em {At. Dat. Nucl. Dat. Tables}\/} {\bf \textbf{34}} 301

\bibitem{Davis1989}
Davis B~F and Chung K~T 1989 {\em Phys. Rev. A\/} {\bf 39} 3942

\bibitem{Gabriel1972}
{Gabriel A H} 1972 {\em {Mon. Not. R. Astr. Soc.}\/} {\bf \textbf{160}} 99

\bibitem{Zhang2013}
{Zhang S B and Yeager D L} 2013 {\em {J. Mol. Struct.}\/} {\bf \textbf{1023}}
  96

\bibitem{Wu1991}
{Wu L and Xi J} 1991 {\em {J. Phys. B: At. Mol. \& Opt. Phys.}\/} {\bf
  \textbf{24}} 3351

\bibitem{Kallman2001}
{Kallman T R and Bautista M A} 2001 {\em {Astrophys. J. Suppl. Ser.}\/} {\bf
  \textbf{134}} 139

\bibitem{Foster2012}
{Foster A R, Ji L, Smith R K and Brickhouse N S} 2012 {\em Astrophys. J\/} {\bf
  \textbf{756}} 128

\end{thebibliography}

\end{document}